\documentclass[]{jfm}

\usepackage[utf8]{inputenc}			% allows direct input of ä,ö,ü,ß
\usepackage[british]{babel}	        % british hyphenation, chapter names, ...
\usepackage{amsmath}				% mathematical environment
\usepackage{amssymb}				% mathematical symbols
\usepackage{mathtools}				% allow use of \mathclap macro
\usepackage{bm}						% allow writing bold mathematical symbols; see also: https://tex.stackexchange.com/questions/3238/bm-package-versus-boldsymbol
\usepackage{graphicx}				% allows including of graphics
\usepackage{xcolor} 				% allows colouring text
\usepackage{comment} 				% allows comments over several lines
\usepackage{array}					% allow, amongst other things, vertical alignment in tables by m{...}
\usepackage{tabularx}				% allow for better tables
\usepackage[english=british]{csquotes} % british style of quotes: enquote, foreignquote, blockquote

\usepackage{hyperref}		% make cross references clickable
\hypersetup{hidelinks} 		% make boxes around links invisible
\hypersetup{
    colorlinks = true,
    urlcolor   = blue,
    citecolor  = black,
}

\usepackage{ragged2e}
\usepackage{newtxtext}
\usepackage{newtxmath}
\usepackage{natbib}
\newcommand{\RomanNumeralCaps}[1]
\linenumbers

\newcommand{\rev}[1]{\textcolor{black}{#1}}

\renewcommand{\Pr}		{\mathrm{Pr}} % overwrite command from amsmath

\renewcommand{\Re}		{\mathrm{Re}} % overwrite command from amsmath
\newcommand{\Ri}		{\mathrm{Ri}}
\newcommand{\Ro}		{\mathrm{Ro}}

\newcommand{\real}   	{\mathfrak{R}} % real part
\newcommand{\dO}		{d_{0}}
\newcommand{\deltaO}	{\delta_{0}}
\newcommand{\UO}		{U_{0}}
\newcommand{\BO}		{B_{0}}
\newcommand{\RO}		{R_{0}}
\newcommand{\ReO}		{\mathrm{Re}_{0}}
\newcommand{\RiO}		{\mathrm{Ri}_{0}}

\newcommand{\Rig}		{\mathrm{Ri}_{\textrm{g}}}
\newcommand{\RigO}		{\mathrm{Ri}_{\textrm{g,0}}}
\newcommand{\Reb}		{\mathrm{Re}_{\textrm{b}}}
\newcommand{\Fr}		{\mathrm{Fr}}
\newcommand{\Rif}		{\mathrm{Ri}_{\textrm{f}}}

\newcommand{\tauadv}   	{\tau_{\textrm{adv}}} % advection time scale
\newcommand{\rhoref}    {\rho_{\textrm{ref}}} % reference density
\newcommand{\pchar}     {p_{\textrm{char}}} % characteristic pressure scale

\newcommand{\tadv}   	{t_{\textrm{adv}}} % advection time scale (non-dimensional)
\newcommand{\tr}   	    {t_{\textrm{r}}} % response time
\newcommand{\tevo}   	{t_{\textrm{evo}}} % evolution or run time of a simulation
\newcommand{\tK}   	    {t_{\textrm{K}}} % Kolmogorov time
\newcommand{\tB}   	    {t_{\textrm{B}}} % Batchelor time
\newcommand{\tS}   	    {t_{\textrm{S}}} % shear time
\newcommand{\tN}   	    {t_{\textrm{N}}} % buoyancy frequency time
\newcommand{\tturb}   	{t_{\textrm{turb}}} % turbulence decay time
\newcommand{\tnu}   	{t_{\nu}} % viscous diffusion time
\newcommand{\tkappa}   	{t_{\kappa}} % thermal diffusion time
\newcommand{\tb}   	    {t_{\textrm{b}}} % buoyancy time

\newcommand{\epsU}      {\varepsilon_{u}} % kinetic energy dissipation rate
\newcommand{\epsB}      {\varepsilon_{b}} % buoyancy variance dissipation rate
\newcommand{\etaK}   	{\eta_{\textrm{K}}} % Kolmogorov scale
\newcommand{\etaO}   	{\eta_{\textrm{O}}} % Ozmidov scale

\newcommand{\Gchi}   	{\Gamma_{\chi}} % flux coefficient based on local dissipation rates
\newcommand{\GchiBar}   {\bar{\Gamma}_{\chi}} % flux coefficient based on averaged dissipation rates

\newcommand{\Gx}   	    {L_{x}} % aspect ratio
\newcommand{\Gy}   	    {L_{y}} % aspect ratio
\newcommand{\Gz}   	    {L_{z}} % aspect ratio
\newcommand{\Gh}   	    {L_{\mathrm{h}}} % aspect ratio
\newcommand{\Ghcrit}   	{L_{\mathrm{h, crit}}} % aspect ratio

\begin{document}

\title{Anisotropy of emergent large-scale dynamics in forced stratified shear flows}

\author{
Philipp P. Vieweg\aff{1}
  \corresp{\email{ppv24@cam.ac.uk}} and
Colm-cille P. Caulfield\aff{2,1}
}
\affiliation{
\aff{1}Department of Applied Mathematics and Theoretical Physics, Wilberforce Road, Cambridge, CB3 0WA, United Kingdom.
\aff{2}Institute for Energy and Environmental Flows, University of Cambridge, Madingley Rise, Madingley Road, Cambridge CB3 0EZ, United Kingdom.
}

\date{\today}

\maketitle

\begin{abstract}
\rev{Although stably stratified shear flows, where the base velocity shear is quasi-continuously forced externally, arise in many geophysically and environmentally relevant circumstances, the emergent dynamics of their ensuing statistically steady stratified turbulence is still an open question. We address this phenomenon in a series of three-dimensional direct numerical simulations using spectral element methods. 
We consider a forced, stably stratified shear flow with an initial bulk Reynolds number $\ReO = 50$, an initial bulk Richardson number $\RiO = 1/80$ (also corresponding to the initial minimum gradient Richardson number $\Rig$), and a fluid of Prandtl number $\Pr = 1$ in horizontally extended domains. 
Although the initial configuration is unstable to a primary Kelvin-Helmholtz instability, the ensuing turbulence is sustained by continuously relaxing the resulting flow back towards the initial profiles of streamwise velocity and buoyancy.
We study statistical as well as structural aspects of the final statistically steady flows, including the flux coefficient $\Gchi$ and dynamically emergent length scales $\Lambda$ associated with the large-scale dynamics, respectively.
Despite the ongoing stirring and mixing, we find that the shear layer half-depth converges to a finite value of $d \approx 8$ (i.e., $\Lambda_{z} \approx 16$) once the horizontal extent of the domain $\Gh \gtrsim 96$. While this implies a final $\Re \approx 400$ and $\Ri \approx 0.1$, we hypothesise that such forced flows \enquote{tune} themselves eventually to a state of a gradient Richardson number $\Rig \lesssim 0.2$, consistently with several previous studies. 
Moreover, provided sufficiently extended domains, we observe the emergence of large-scale flow structures with spanwise $\Lambda_{y} \approx 50$ and streamwise $\Lambda_{x} \lesssim 115$. 
Clearly, these observations demonstrate the marked anisotropy of characteristic emergent length scales, even for such \enquote{weakly stratified} forced shear flows. We conjecture that the actual emergent streamwise structures are a vestigial \enquote{imprint} in the sheared turbulent flow of the primary linear instability of the converged deepened turbulent shear layer.}
\end{abstract}
\keywords{forced stratified shear flow, emergent phenomena, anisotropy, pattern formation, direct numerical simulation}

%------------------------------------------------------------------------------------------
\section{Introduction}
\label{sec:Introduction}

Stably stratified (vertical) shear flows, where both the background fluid buoyancy (i.e. the appropriately scaled negative density perturbation) and horizontal flow velocity vary with height, are ubiquitous in geophysical contexts. There has been a very large body of work considering the ways in which such flows behave, (as evidenced by a sequence of reviews such as \cite{fernando_1991,peltier_2003,ivey_2008,caulfield_2020} and \cite{caulfield_2021}) with significant focus on the (possible) existence of turbulence extracting and converting the kinetic energy in the background shear and the associated enhanced (irreversible) turbulent mixing. Understanding (and parameterising) such turbulent mixing of stratified fluids is a key challenge in geophysical flows, as the transport of momentum, heat and other scalars (such as dissolved gases, pollution and microplastics for example) is both a crucial process and a phenomenon of great uncertainty for the description of the world's climate and environment (see for example the reviews of \cite{wunsch_2004,wunsch_2009,gregg_2018}). 

%------------------------------------------------------------------------------------------
\begin{figure}
\centering
\includegraphics[width = \textwidth]{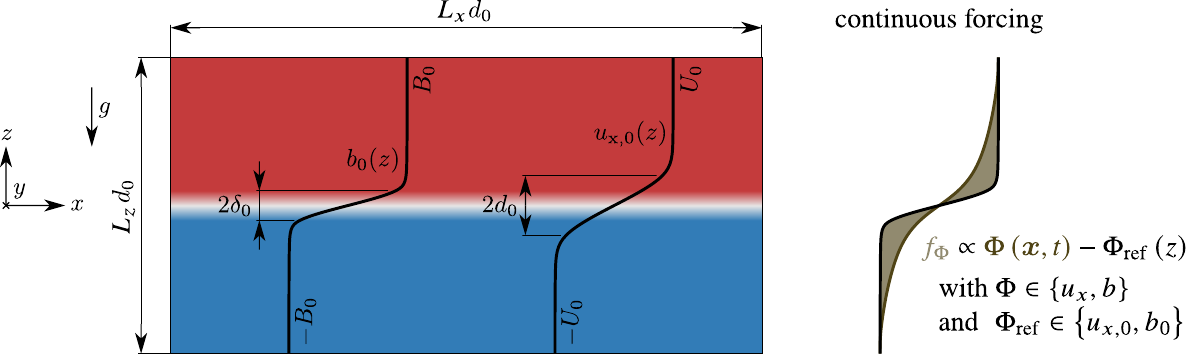}
\caption{\justifying{Configuration of the forced stratified shear flows considered here. 
While the initial buoyancy profile $b_{0}$ is statically stable, the imposed initial velocity profile $u_{x, 0}$ may induce flow instabilities that lead to a transition to turbulence. 
A continuous forcing $f_{\Phi}$ may inject the required energy to sustain this turbulence, ensuring statistically stationary dynamics.
}}
\label{fig:schematic_configuration}
\end{figure}
%------------------------------------------------------------------------------------------

In essence,  as shown schematically in figure \ref{fig:schematic_configuration}, stably stratified shear flows are characterised by a competition between a stabilising buoyancy and a de-stabilising velocity (or shear) profile. 
However, gaining an understanding of fundamental aspects of this deceptively simple set-up continues to prove exceptionally complicated. 
Considering initial value problems of (initially) laminar velocity and density profiles, it is well-known that such flows can be prone to a range of primary flow instabilities (see the review of \cite{caulfield_2021}) that effectively rearrange the strip of spanwise vorticity into trains of elliptical vortices (such as the classic Kelvin-Helmholtz \enquote{billows}, i.e. the saturated manifestation of the Kelvin-Helmholtz Instability (KHI)). As these vortices themselves are prone (at least for sufficiently high Reynolds number) to a \enquote{zoo} of secondary instabilities (a nomenclature proposed by \cite{mashayek_2012a,mashayek_2012b}), turbulence transition is triggered and mixing as well as dissipation are eventually significantly enhanced.

Such mixing events can extend for a significant length of time -- particularly when the flow is prone primarily to the so-called \enquote{Holmboe Wave Instability} (HWI) \citep{Holmboe62}, as discussed in detail by \cite{salehipour_2016,salehipour_2018} -- \rev{and can also be accompanied by the formation of larger-scale emergent structures, especially in sufficiently spatially extended flow domains \citep{Watanabe2021}}. 
The particular life cycle of such mixing events can be strongly sensitive to initial conditions, as demonstrated conclusively by \cite{liu_2022}, due to competition between different members of the secondary instability \enquote{zoo}, especially involving the subharmonic \enquote{pairing} instability of neighbouring primary Kelvin-Helmholtz billows. As demonstrated by \cite{mashayek_2013}, sufficiently strong stratification can disrupt and hence suppress such pairing (or more precisely \enquote{merging}) events due to the energetic costs of the inherent vertical motions. 
\rev{Given only finite computational resources}, such suppression is often cited as a reason to restrict \rev{(the inevitably finite)} domains to one (or at most two) wavelengths of the primary instabilities, allowing to reach higher Reynolds numbers.

However, the horizontal extent of the \rev{(numerical or experimental)} domain may \rev{clearly} affect the flow dynamics. As shown by \cite{scinocca_2000}, rich dynamics can occur in longer streamwise domains, where primary instabilities with close wavelengths (and hence close linear growth rates) can compete as they grow to finite amplitude if the (conventional) imposition of periodic streamwise boundary conditions does not quantise the accessible wavelengths of possible instabilities too severely. 
Analogous issues also arise in the spanwise direction. Many studies consider relatively narrow domains, in the sense that the spanwise extent is (often significantly) smaller than the characteristic (streamwise) wavelength of the primary instability. Such domains allow many of the (essentially local) secondary instabilities to develop and hence trigger turbulence transition. However, as can be straightforwardly observed in sufficiently wide tilting tank experiments \citep{thorpe_1985,thorpe_1987,caulfield_1996,thorpe_2002} and sufficiently wide computational domains, as clearly demonstrated by \cite{fritts_2022a,fritts_2022b}, inherently three-dimensional \enquote{knots}, \enquote{tubes}, and billow \enquote{defects} can develop in the spanwise direction on scales of the order of (but typically larger than) the primary instability's  most unstable streamwise wavelength. 

Moreover, there will always be an inevitable decay of these reported flow structures unless the underlying driving mechanism (i.e. the shear) is replenished. 
Although such initial-value-problem mixing events are undoubtedly of geophysical interest and indeed demonstrate the formation of transient larger-scale dynamics \rev{exhibiting the so-called \enquote{lifecycle} of a shear-driven mixing event}, it is important to remember that such initial-value-problem mixing events are really just one end member class of the possible flows of geophysical interest.

The other obvious end member class is the class of continuously forced flows, where the \enquote{background} velocity and density (or equivalently buoyancy) profiles are driven by some external, quasi-steady forcing. Possible candidate mechanisms for such geophysically relevant forcings include wind, solar radiation and resulting evaporation at the surface of the ocean \citep{Thorpe2005}, tidal forcing  \citep{StLaurent2002}, active matter living in water \citep{Castro2022}, or continuous outflows from rivers \citep{Uncles2011}. While this list is not meant to be complete in any sense, it illustrates that quasi-steady forcings do occur in geophysically relevant situations. An (artificial) volumetric forcing is a particularly convenient way to mimic these natural complex mechanisms in a simplified manner. As visualised on the right side of figure \ref{fig:schematic_configuration}, such a forcing may be defined to relax the \rev{time-dependent} local profiles. 

Such a forcing was \rev{introduced} by \citet{Smith2021}, who demonstrated that -- after an initial transient where primary instabilities (either KHI or HWI) develop and (inevitably) break down -- the ensuing turbulence could be sustained over arbitrarily long times, thus enabling statistically steady statistics of the flow to be calculated. There is a clear attraction to considering such forced flows due principally to the inherent removal of the confounding effects of the life cycle of the mixing event from the turbulence statistics. 
Therefore, such flows seem the natural test-bed to \rev{study the emergent and sustained turbulent self-organisation of the flow on large scales. 
This endeavour, however, comes necessarily with the challenge of first diagnosing whether or not there is actually an \textit{inherent} convergence of the resulting depth of the turbulent shear layer, and second identifying its minimal required extent to enable such statistical convergence (that is unbiased by the extent of the domain).}
  
\rev{Given that \cite{Smith2021}}, similar to the initial value problem flows discussed above, also considered relatively small computational domains only, it is not at all clear whether or not their statistically stationary flow remained unaffected by the computational domain size.
An equivalent question to ask is \rev{\enquote{what are the unconstrained emergent length scales of a forced stratified shear flow?} Answering that (open) question is the central objective of this paper.}

To address this \rev{objective}, the rest of the paper is organised as follows. 
In section \ref{sec:Numerical_approach} we present our numerical approach before we list the range of considered \rev{control parameters and} computational domains at the start of section \ref{sec:Results}. For simplicity, we focus on \rev{one set of parameters that enables} a relatively weakly stratified flow which is prone to primary KHI. \rev{After briefly identifying} some interesting early-time dynamics that may possibly be affected by \rev{molecular} diffusion \rev{due to the choice of control parameters} \rev{(but is not of principal interest here), we shift our focus to the statistically steady turbulent state that is sustained by the volumetric forcing.}
We identify \rev{emergent,} strongly anisotropic length scales which \rev{are proven to converge} in sufficiently extended domains. We demonstrate conclusively that key statistics of the flow, including in particular those related to irreversible mixing, are sensitive to the \rev{large-scale self-organisation of the flow or equivalently the} extent of the computational domain. \rev{From a practical perspective, we} remark that for convergent statistics, perhaps surprisingly, the flow domain may need to be extraordinarily extended (i.e. of order $100$ times larger) compared to the initial shear-layer half-depth. 
\rev{We discuss the implications of the discovered emergent large-scale dynamics with an emphasis on the flow physics in section \ref{sec:Discussion} and propose their dynamical origin. Finally, we draw our conclusions in section \ref{sec:Conclusions}.}

%------------------------------------------------------------------------------------------
\section{Numerical approach}
\label{sec:Numerical_approach}

\subsection{Governing equations}
\label{subsec:Governing_equations}

We consider an incompressible flow based on the Oberbeck-Boussinesq approximation with a linear equation of state. 
%citep{Oberbeck1879, Boussinesq1903}. 
The three-dimensional equations of motion are non-dimensionalised using the (dimensional) initial magnitudes of the streamwise velocity $\UO$ and buoyancy $\BO$, as well as the shear layer half-depth $\dO$ as shown in figure \ref{fig:schematic_configuration}. 
Using the advective time scale $\tauadv := \dO / \UO$ and the appropriate characteristic pressure scale $\pchar := \UO^{2} \rhoref$, non-dimensional variables (marked here with a tilde) can be related to the dimensional variables as follows:
\begin{equation}
\label{eq:def_non-dimensionalisation}
\bm{x} = \dO               \tilde{\bm{x}} , \qquad 
\bm{u} = \UO               \tilde{\bm{u}} , \qquad 
b      = \BO               \tilde{    b } , \qquad 
t      = \tauadv           \tilde{    t } , \qquad 
p      = p_{\textrm{char}} \tilde{    p }
\end{equation}
Henceforth, we focus on non-dimensional variables, and so drop the tildes from all variables.

As a consequence, the non-dimensional governing equations are
\begin{align}
\label{eq:CE}
\nabla \cdot \bm{u} &= 0 , \\
\label{eq:NSE}
\frac{\partial \bm{u}}{\partial t} + \left( \bm{u} \cdot \nabla \right) \bm{u} &= - \nabla p + \frac{1}{\ReO} \nabla^{2} \bm{u} + \RiO b \bm{e}_{z} + f_{u} \bm{e}_{x} , \\
\label{eq:BE}
\frac{\partial b}{\partial t} + \left( \bm{u} \cdot \nabla \right) b &= \frac{1}{\ReO \Pr} \nabla^{2} b + f_{b} ,
\end{align}
where $\bm{u}$, $b$ and $p$ represent the velocity, buoyancy, and modified pressure field, respectively. The precise form of the volumetric forcing terms $f_{u}$ and $f_{b}$ will be defined below. The buoyancy $b := - \rho^{'} g / \rhoref$ and corresponds to the negative of the reduced gravity, so $\rho'$ is the deviation from the reference density $\rhoref$. Three non-dimensional parameters naturally emerge from this scaling: the Prandtl number, the initial bulk Reynolds number, and the initial bulk Richardson number,
\begin{equation}
\label{eq:def_Pr_Re0_Ri0}
\Pr  := \frac{\nu}{\kappa} , \qquad
\ReO := \frac{\UO \dO}{\nu} , \qquad
\RiO := \frac{\BO \dO}{\UO^{2}} ,
\end{equation}
respectively, where $\nu$ is the kinematic viscosity and $\kappa$ is the buoyancy diffusivity. We restrict attention to statically stable situations where $\RiO > 0$.

The volumetric forcing terms $f_{u}$ and $f_{b}$ in equations \eqref{eq:NSE} and \eqref{eq:BE} are a crucial aspect of our configuration. Following \citet{Smith2021}, we consider
\begin{alignat}{8}
\label{eq:def_fu}
f_{u} &:= - \frac{1}{\tr} \left[ u_{x} - u_{x, 0} \right] &
\qquad &\textrm{with} \quad &
u_{x, 0} \left( z \right) &:= \tanh \left( z \right) , \\
\label{eq:def_fb}
f_{b} &:= - \frac{1}{\tr} \left[ b - b_{0} \right] &
\qquad &\textrm{with} \quad&
b_{0} \left( z \right) &:= \tanh \left( \RO z \right) 
\end{alignat}
where $\tr$ is the response time while $u_{x, 0}$ and $b_{0}$ represent the initial streamwise velocity and buoyancy base profiles to which the flow \rev{relaxes back under the effect of the forcing, at least in principle}. In this context, $\RO := \dO / \deltaO$ defines the ratio of initial interface half-depths (i.e. $\deltaO$ represents the dimensional initial buoyancy interface half-depth) with $\RO = \sqrt{\Pr}$ following the diffusive arguments presented by \cite{smyth_1988}. 
In essence, these forcing terms are idealized models of geophysically relevant processes \rev{(see again our introduction)} that tend to restore the initial profiles $u_{x, 0}$ and $b_{0}$, and thus sustain turbulence over arbitrarily long times. 
\rev{In this context, we stress that the response time $\tr$ quantifies the relative strength of these superposed processes. While $\tr \rightarrow 0$ implies vigorously enforcing the sharp initial profiles, $\tr \rightarrow \infty$ corresponds to  the disappearance of the forcing. As shown by \citet{Smith2021}, the broad sweetspot -- to allow for sustained turbulence without prescribing a significant signal of the forcing -- lies in between and tends to preserve the primary instability associated with the underlying base profiles. Finally, we remark that, although the included base profiles are time-independent, the resulting rate of momentum or buoyancy injection is typically not constant in space or time.}

Hence, the governing equations \eqref{eq:CE} -- \eqref{eq:BE} are fully specified via four control parameters: $\Pr$, $\ReO$, $\RiO$, and $\tr$.

\subsection{Domain, boundary and initial conditions, and numerical code}
\label{subsec:domain_boundary_and_initial_conditions_and_numerical_code}

As indicated by figure \ref{fig:schematic_configuration}, \rev{the volume $V := \Gx \times \Gy \times \Gz$ of the numerical domain, i.e. the product of the} streamwise, spanwise, and vertical extents $\Gx$, $\Gy$, and $\Gz$, respectively. Both the midpoint of the shear layer and the midpoint of the buoyancy distribution are located at midplane, $z = 0$, with a horizontal cross-section $A := \Gx \times \Gy$.
We consider a horizontally periodic domain where any quantity
\begin{equation}
\label{eq:BC_periodic}
\Phi \left( \bm{x} \right) = \Phi \left( \bm{x} + i_x \Gx \bm{e}_{x} + i_y \Gy \bm{e}_{y} \right) 
\qquad \textrm{given} \quad i_{x,y} \in \mathbb{N} .
\end{equation}
Additionally, we apply free-slip and no-flux boundary conditions at the top and bottom, so
\begin{equation}
\label{eq:BC_free-slip}
u_{z} = \frac{\partial u_{x}}{\partial z} = \frac{\partial u_{y}}{\partial z} = 0 
\qquad \textrm{and} \qquad 
\frac{\partial b}{\partial z} = 0 
\qquad \textrm{at} \quad 
z = \pm \frac{\Gz}{2} .
\end{equation}
Our initial condition is given by 
\begin{equation}
\label{eq:IC}
u_{x} = u_{x, 0} , \qquad 
u_{y} = u_{z} = 0 , \qquad 
\textrm{and} \qquad 
b = b_{0} + \Upsilon 
\qquad \textrm{at} \quad 
t = 0 .
\end{equation}
The random fluctuations $- 10^{-3} \leq \Upsilon \left( \bm{x} \right) \leq 10^{-3}$ \enquote{seed} the development of primary instabilities. 

We solve the coupled governing equations \eqref{eq:CE} -- \eqref{eq:BE}, subject to these boundary and initial conditions \eqref{eq:BC_periodic} -- \eqref{eq:IC}, using the GPU-accelerated spectral element solver NekRS \citep{Fischer1997,Scheel2013, Fischer2022}. 
\rev{As the spatial resolution of spectral element methods is determined by the combination of the number of spectral elements $N_{\textrm{e}} = N_{\mathrm{e}, x} \times N_{\mathrm{e}, y} \times N_{\mathrm{e}, z}$ and polynomial order $N$ of the spectral expansion within each element \citep{Deville2002, Vieweg2023a}, }
such methods can -- as shown in more detail in appendix \ref{app:resolving_shear_flows} -- accommodate different required spatial resolutions across the domain and are thus well suited to resolve shear flows efficiently. 
This is particularly important given degrees of freedom of up to $N_{\textrm{dof}} \approx 3 N_{\textrm{e}} N^{3} \sim \mathcal{O} \left( 10^{9} \right)$ \rev{and the lengthy flow evolution} in our present study.

%---------------------------------------------------------------------------
\newcommand{\hp}{\hphantom{00}}
\newcommand{\hpp}{\hphantom{0}}
\begin{table}
\centering
\begin{tabular}{@{\hskip 0mm} r @{\hskip 3.5mm} r @{\hskip 3.5mm} c @{\hskip 3.5mm} c @{\hskip 3.5mm} c @{\hskip 3.5mm} c @{\hskip 3.5mm} c @{\hskip 0mm}}
$\Gx \times \Gy \thickspace \thickspace$ & $N_{\mathrm{e}, x} \times N_{\mathrm{e}, y}$ & $d$ & $\delta$ & $R$ & $\Re$ & $\Ri$ \\
% This was the header - here comes the content.
  $16^{2}$ \hp     &    $9^{2}$ \hp     & $2.66 \pm 0.16$ & $2.80 \pm 0.14$ & $0.95 \pm 0.04$ & $133 \pm \hpp 8$ & $0.0333 \pm 0.0020$ \\
  $32^{2}$ \hp     &   $18^{2}$ \hp     & $5.35 \pm 0.34$ & $5.54 \pm 0.35$ & $0.96 \pm 0.02$ & $267 \pm     17$ & $0.0668 \pm 0.0043$ \\
  $64^{2}$ \hp     &   $36^{2}$ \hp     & $7.55 \pm 0.47$ & $7.75 \pm 0.36$ & $0.97 \pm 0.03$ & $377 \pm     24$ & $0.0943 \pm 0.0059$ \\
  $96^{2}$ \hp     &   $54^{2}$ \hp     & $7.92 \pm 0.53$ & $8.27 \pm 0.56$ & $0.96 \pm 0.02$ & $396 \pm     26$ & $0.0990 \pm 0.0066$ \\
 $128^{2}$ \hp     &   $72^{2}$ \hp     & $8.05 \pm 0.42$ & $8.41 \pm 0.45$ & $0.96 \pm 0.01$ & $403 \pm     21$ & $0.1007 \pm 0.0053$ \\
 $256^{2}$ \hp     &  $144^{2}$ \hp     & $7.90 \pm 0.18$ & $8.22 \pm 0.17$ & $0.96 \pm 0.01$ & $395 \pm \hpp 9$ & $0.0987 \pm 0.0022$ \\
 $512^{2}$ \hp     &  $288^{2}$ \hp     & $7.73 \pm 0.08$ & $8.06 \pm 0.07$ & $0.96 \pm 0.00$ & $387 \pm \hpp 4$ & $0.0967 \pm 0.0010$ \\
 \\
$2048 \times  512$ & $1152 \times 288$  & $7.79 \pm 0.03$ & $8.12 \pm 0.03$ & $0.96 \pm 0.00$ & $390 \pm \hpp 1$ & $0.0974 \pm 0.0003$ \\
 $512 \times 2048$ &  $288 \times 1152$ & $7.77 \pm 0.04$ & $8.10 \pm 0.04$ & $0.96 \pm 0.00$ & $388 \pm \hpp 2$ & $0.0971 \pm 0.0005$
\end{tabular}
\caption{\justifying{Simulation parameters.
The Prandtl number $\Pr = 1$, initial bulk Reynolds number $\ReO = 50$, initial bulk Richardson number $\RiO = 0.0125$, response time $\tr = 100$, and initial ratio of interface half-depths $\RO = 1$ in a horizontally periodic domain. In the vertical direction, the domain has an aspect ratio $\Gz = 48$ spanned by $N_{\mathrm{e}, z} = 18$ non-uniformly distributed spectral elements (see appendix \ref{app:resolving_shear_flows} for more details) together with free-slip and no-flux boundary conditions for the velocity and buoyancy field, respectively. The polynomial order $N = 6$. Although the total evolution or run time of each flow $\tevo = 5,040$, this work focuses on the statistically stationary dynamics during the last $\Delta t = 3,000$ only.
For each simulation, this table lists the horizontal \rev{cross-sectional areas} $\Gx \times \Gy$ and corresponding numbers of uniformly distributed spectral elements $N_{\mathrm{e}, x} \times N_{\mathrm{e}, y}$. 
Moreover, we include the final \rev{emergent} shear layer half-depth $d$ of the streamwise velocity field as well as the buoyancy interface half-depth $\delta$, their ratio $R$, the bulk Reynolds number $\Re$, as well as the bulk Richardson number $\Ri$, \rev{listing} both their temporal means and standard deviations.
}}
\label{tab:simulation_parameters}
\end{table}
\let\hp\undefined
\let\hpp\undefined
%--------------------------------------------------------------------------

%------------------------------------------------------------------------------------------
\section{Results}
\label{sec:Results}

This study considers forced stratified shear flows at $\Pr = 1$, $\ReO = 50$, $\RiO = 0.0125$, and $\tr = 100$ in domains of $\Gz = 48$. 
As will be shown in more detail in \rev{section \ref{subsec:Interaction_between_the_two_layers_of_fluid}}, these parameters are associated with the minimum value of the initial gradient Richardson number $\RigO \left( z = 0 \right) = \RiO \RO$ which is sufficiently small to enable the development of primary KHI. 
We investigate and quantify the emergent dynamics of the flow while varying the horizontal extents of the domain, $\Gx$ and $\Gy$. Normally, we consider domains of \textit{square} horizontal cross-section with $\Gh \equiv \Gx = \Gy$ ranging from $16$ to $512$. Table \ref{tab:simulation_parameters} summarises all our considered domains.

\subsection{Typical evolution of a forced stratified shear flow}
\label{subsec:Exemplary_evolution_of_a_forced_stratified_shear_flow}

%------------------------------------------------------------------------------------------
\begin{figure}
\centering
\includegraphics[scale = 1.0]{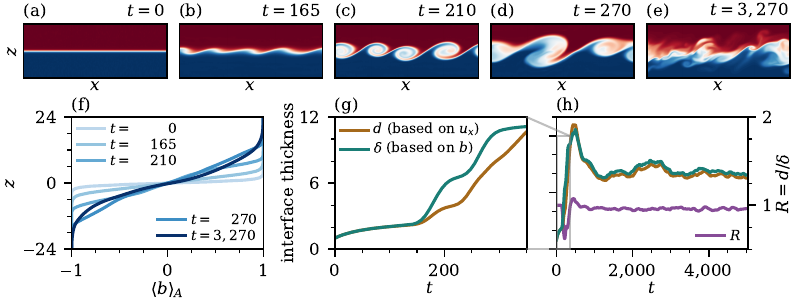}
\caption{\justifying{Temporal evolution of forced, stratified shear flows.
(a -- e) The flow is prone to a primary KHI, leading eventually to \enquote{overturning billows} and streamwise mergers. A continuous forcing sustains the induced turbulence for arbitrarily long times. 
During this evolution of the flow, (f -- h) the interface broadens before reaching a statistically stationary depth. 
Note that for this relatively small $\ReO$, as shown in (g), \rev{molecular} diffusion of the shear layer and density interface dominates the development of the primary instability until the depths of the shear layer and density interface have approximately doubled.}
In this figure, $\Gh = 128$ while panels (a -- e) visualise $b \left( x, y = 0, z, t \right)$ with the colour bar matching figure \ref{fig:emerging_horizontally_extended_dynamics} (l, p).
}
\label{fig:temporal_evolution}
\end{figure}
%------------------------------------------------------------------------------------------

We first consider the evolution of a typical flow in a domain with  horizontal extent of $\Gh = 128$. 
Figures \ref{fig:temporal_evolution} (a -- e) show snapshots of the  instantaneous buoyancy field $b \left( x, y = 0, z, t \right)$ in vertical slices normal to the spanwise direction.  At early times, the flow is prone to a primary Kelvin-Helmholtz instability which, driven by the vertical shear, develops into a train of KH billows (panel (c)) that merge subsequently (panel (d)). These primary billows break down, and the buoyancy interface broadens (panel (f)). \rev{Note that we use $\langle f \rangle_{\Phi}$ to denote the average value of some flow variable $f$ computed over $\Phi$, where the averaging domain $\Phi$ may be a combination of spatial and temporal dimensions. Here, time-averages cover the flows' last $\Delta t = 3,000$, i.e. the time interval associated with the statistically stationary regime of primary interest.}
 
If this was an initial value problem, turbulence would die out shortly after $t=270$ due to the combined (and inter-related) effects of 
enhanced dissipation and broadening of both the shear layer as well as the density interface, eventually leading to an increased (and, according to the so-called Miles-Howard criterion \citep{Miles1961,Howard1961}, indeed linearly stable) minimum gradient Richardson number. 
However, volumetric forcing sustains the induced turbulence over arbitrarily long times. This is underlined by panel (e), which shows a snapshot during this late statistically steady turbulent state of the flow. Here, we focus largely on this late, statistically stationary dynamics. 

Across the evolution of the flow, we quantify the associated broadening of the shear layer and buoyancy interface via
\begin{equation}
\label{eq:def_interface_thicknesses}
d \left( t \right) := \frac{1}{2} \int_{- \Gz / 2}^{+ \Gz / 2} \left( 1 - \langle u_{x} \rangle_{A}^{2} \right) dz
\qquad \textrm{and} \qquad 
\delta \left( t \right) := \frac{1}{2} \int_{- \Gz / 2}^{+ \Gz / 2} \left( 1 - \langle b \rangle_{A}^{2} \right) dz ,
\end{equation}
respectively, using the approach proposed by \citet{Smyth2000}. Together with the initial profiles defined in equations \eqref{eq:def_fu} and \eqref{eq:def_fb}, these functions yield $d \left( t = 0 \right) = 1$ and $\delta \left( t = 0 \right) = 1 / \RO$ (as these non-dimensional lengths $\left\{ d, \delta \right\}$ are measured in units of $\dO$, see again equation \eqref{eq:def_non-dimensionalisation}). Similarly, the time-dependent ratio of interface depths $R \left( t \right) := d / \delta$ with $R \left( t = 0 \right) = \RO$.
Figures \ref{fig:temporal_evolution} (g, h) highlight that both $d$ and $\delta$ converge after an initial transient \rev{of length $\lesssim 10^{3} \tauadv$} to statistically stationary values $\left\{ d, \delta \right\} \gg 1$. In other words, the interfaces have deepened significantly \rev{and, remarkably, reached a finite value characterised by a balance between the ongoing forcing (attempting to relax the interfaces
back to their original depths) and mixing (tending to deepen the interfaces). We establish the independence of this finite value from the extent of the domain in the subsequent sub-sections and hypothesise a physical reason for this convergence in section \ref{sec:Discussion}.} Consistently with the fact that the flow remains \enquote{weakly} stratified, the ratio $R \approx 1$ at late times. Here, \enquote{weakly stratified} is meant in the specific sense that the turbulent diffusivity of buoyancy closely follows the turbulent diffusivity of the momentum. Equivalently, the turbulent Prandtl number is close to one, and so the buoyancy field is at least in some sense slaved to the velocity field and behaves like a passive scalar \rev{\citep{turner_1973}. 
Due to the particular form of the forcing, with shear continually being reinforced, it is also appropriate to consider this flow to be \enquote{shear-dominated} within the nomenclature of \cite{Mater2014}.}

We note in passing that, as highlighted in panel (g), the onset of the primary KHI (evidenced by the significant change in the rate of the growth of $d$ and $\delta$ around $t\simeq 150$) \rev{appears} only after a period of significant diffusive deepening of both the shear layer and the buoyancy interface. As the shear layer half-depth $d$ approximately doubles \rev{during this initial period of time}, the onset of the primary instability \rev{triggers} a significantly larger wavelength. \rev{In other words, the characteristic scales of the  delayed KHI are set relative to $d$ rather than $\dO$.}
An investigation of this interesting early-time interaction between diffusion and instability onset is beyond the scope of this paper but undoubtedly worthy of further, more detailed consideration. 

\rev{Having reached the final, statistically stationary state,} the above definition of $d \left ( t \right)$ enables the computation of associated time-dependent values of the bulk Reynolds and Richardson number via
\begin{equation}
\label{eq:def_final_Re_Ri}
\Re \left( t \right) := \frac{\UO \left( d \thickspace \dO \right)}{\nu} = \ReO \thickspace d 
\qquad \textrm{and} \qquad 
\Ri \left( t \right) := \frac{\BO \left( d \thickspace \dO \right)}{\UO^{2}} = \RiO \thickspace d 
\end{equation}
with, again, $\Re \left( t = 0 \right) = \ReO$ and $\Ri \left( t = 0 \right) = \RiO$. We stress that, since $\langle d \rangle_{t} \approx 8 \gg 1$ at our late times of interest, the final flow has an effective temporal average $\langle \Re \rangle_{t} \approx 400 \gg \ReO$ and has, thus, become much more turbulent than what one might have expected from the relatively small initial value of $\ReO=50$.

Table \ref{tab:simulation_parameters} lists the temporal averages and standard deviations of $d$, $\delta$, $R$, $\Re$, and $\Ri$ during the late statistically stationary dynamics of all our simulations. We discuss their trends with respect to the horizontal domain size $\Gh$ in section \ref{subsec:Convergence_of_vertical_stirring_and_dissipation_for_extended_domains}.

\subsection{\rev{Sustained turbulent interactions between the two layers}}
\label{subsec:Interaction_between_the_two_layers_of_fluid}

%------------------------------------------------------------------------------------------
\begin{figure}
\centering
\includegraphics[scale = 1.0]{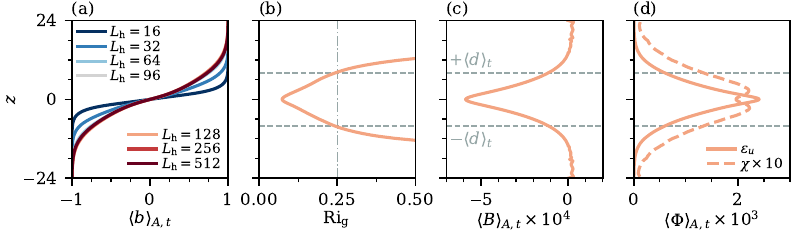}
\caption{\justifying{Sustained turbulent interactions.
(a) Despite statistical stationarity, the deepening of the density interface can be affected by the horizontal extent of the domain $\Gh \equiv \Gx = \Gy$ before it converges eventually. 
\rev{This deepening tends to stabilise the emergent flow, resulting in} a significantly increased minimum value of $0.074$ of (b) the average late $\Rig$. Nevertheless, we find $\Rig \leq 0.25$ (with this canonical value being marked with a vertical line) throughout the turbulent \enquote{mixing zone}.
The resulting associated mixing in this zone is underlined by high amplitudes in 
(c) the stabilising vertical buoyancy advection (i.e. the buoyancy flux) $B$ and  
(d) the dissipation rates of kinetic energy and scaled buoyancy variance, $\epsU$ and $\chi$.
}}
\label{fig:interaction_of_the_two_layers}
\end{figure}
%------------------------------------------------------------------------------------------

\rev{Even though a continuous forcing allows the existence of a  statistically stationary regime, the associated sustained turbulent interactions between the fluid regions above and below the shear layer may yet be impacted by the extent of the domain.}
In figure \ref{fig:interaction_of_the_two_layers} (a) we plot the mean vertical buoyancy profiles $\langle b \rangle_{A, t}$ associated with the late-time dynamics \rev{of the forced stratified shear flow}. We find that these profiles depend strongly on the horizontal extent $\Gh$ of the domain. Although $\Gh = 16 \gg 1$ even for our smallest domain, the interface deepens for increasing $\Gh$ and a convergence seems to be reached for $\Gh \gtrsim 64$ only. This suggests that the vertical \rev{stirring and resultant} mixing of buoyancy strongly depends on the horizontal extent of the domain. 

This observed deepening of the buoyancy interface is accompanied by a deepening of the shear layer, both affecting in turn the gradient Richardson number
\begin{equation} 
\label{eq:def_gradient_Richardson_number}
\Rig \left( z \right) := \frac{\langle N^{2} \rangle_{A, t}}{\langle S \rangle_{A, t}^{2}} \equiv \left( \frac{\tS}{\tN} \right)^{2}
\end{equation}
via the buoyancy frequency
(for which averaging is typically associated with $\partial b / \partial z$)
and vertical shear, 
\begin{equation}
\label{eq:def_buoyancy_frequency}
N := \sqrt{\RiO \frac {\partial b}{\partial z}} 
\qquad \textrm{and} \qquad
S := \frac{\partial u_{x}}{\partial z} ,
\end{equation}
respectively. 
\rev{We remark that it is also possible to associate the initial profiles from equations \eqref{eq:def_fu} and \eqref{eq:def_fb} with an initial gradient Richardson number $\Rig \left( z, t = 0 \right) = \RigO := \RiO \left( \partial b_{0} / \partial z \right) / \left( \partial u_{x,0} / \partial z \right )^{2}$.}
As shown in figure \ref{fig:interaction_of_the_two_layers} (b), the minimum value of $\Rig$ has significantly increased \rev{from $\RigO \left( z = 0 \right) = 0.0125$} to $\Rig \left( z = 0 \right) \approx 0.074$ provided $\Gh = 128$. Note that, due to the square in the denominator of the definition, the reduction of shear overpowers the simultaneous reduction of buoyancy stratification.
\rev{As indicated on the right of equation \eqref{eq:def_gradient_Richardson_number}, $\Rig$ can also be interpreted as a ratio of time scales associated with the average shear and stratification, an observation we will discuss in detail in section \ref{sec:Discussion}.}

From this panel (b), it is also apparent that $\Rig \leq 0.25$ within the region $ - \langle d \rangle_{t} \leq z \leq \langle d \rangle_{t}$. Although the flow is undoubtedly turbulent -- and so the Miles-Howard criterion \citep{Miles1961,Howard1961} of (steady) linear inviscid stability theory 
is definitely not applicable --, such relatively small values of $\Rig$ are necessary for turbulence to survive and so we refer to this zone as the  turbulent \textit{mixing zone}. 

Indeed, this nomenclature is also strongly justified (as shown in figures \ref{fig:interaction_of_the_two_layers} (c, d)) by 
considering the vertical buoyancy advection (frequently called the buoyancy flux)
\begin{equation}
\label{eq:def_vertical_buoyancy_advection}
B := \RiO u_{z} b 
\end{equation}
as well as the dissipation rates of kinetic energy $\epsU$ and scaled buoyancy variance $\chi$
\begin{alignat}{5}
\label{eq:def_dissipation_rate_kinetic_energy}
\epsU &:= \frac{2}{\ReO} \bm{S}^{2} 
\qquad &&\textrm{with} \quad
\bm{S} &&:= \frac{1}{2} \left[ \left( \nabla \bm{u} \right) + \left( \nabla \bm{u} \right)^{T} \right] 
\quad \textrm{and} \\
\label{eq:def_dissipation_rate_buoyancy_variance}
\chi &:= \frac{\RiO^{2} \thickspace \epsB}{\langle N^{2} \rangle_{A, t}}
\qquad &&\textrm{with} \quad
\epsB &&:= \frac{1}{\ReO \Pr} \left( \nabla b \right)^{2} ,
\end{alignat}
respectively,
which naturally emerge in the evolution equations for kinetic energy and scaled buoyancy variance (see for example the review of \cite{caulfield_2021} for a derivation). 
Here, $\bm{S}$ represents the strain rate tensor and $\epsB$ the (unscaled) buoyancy variance dissipation rate. Note that the scaling via $\langle N^{2} \rangle_{A, t}$ results in identical physical units for the associated dimensional dissipation rates. 
We emphasise this point by the comparison
\begin{alignat}{3}
\label{eq:eps_u_scale}
\epsU &= \frac{\UO^{3}}{ \dO } \thickspace \tilde{\varepsilon}_{u}, && &&\\
\label{eq:eps_b_scale}
\epsB &= \frac{\BO^{2} \UO}{\dO} \thickspace \tilde{\varepsilon}_{b} &&= \frac{\UO^{5}}{\dO^{3}} \RiO^{2} \thickspace \tilde{\varepsilon}_{b}, &&\\ 
\label{eq:chi_scale} 
\chi &=  \frac{\BO^{2} \dO}{\UO} \thickspace \tilde{\tilde{\chi}} &&= \frac{\UO^{3}}{\dO} \RiO^{2} \thickspace \tilde{\tilde{\chi}} &&= \frac{\UO^{3}}{\dO} \thickspace \tilde{\chi}, 
\end{alignat}
where we have, for improved clarity, re-introduced tildes for non-dimensional quantities in the above three equations only.
This property of identical physical units is particularly helpful when studying the exchange of kinetic and potential energy, as buoyancy variance is closely related to the concept of \enquote{available potential energy} \rev{as originally introduced by \cite{lorenz_1955} and further developed by \cite{Winters95}, see for example the review of \cite{caulfield_2021} for further discussion}.

As shown in panels (c -- d) of figure \ref{fig:interaction_of_the_two_layers}, all of these quantities exhibit pronounced peaks close to the midplane. 
However, while the vertical buoyancy advection $B$ introduces a macroscopic stirring that is generally reversible, it leads to irreversible dissipation via both $\epsU$ and $\epsB$ due to the inherent coupling of $\bm{u}$ and $b$. 
As $\langle B \rangle_{V, t} < 0$, this stirring comes with an overall stabilising effect on the configuration. Furthermore, although enhanced values of $\epsU$ and $\chi$ extend beyond the mixing zone, it is clear that \rev{this mixing zone} contains the vast majority of the enhanced dissipation in \rev{this regime}. 

\rev{Hence, the emergent turbulent shear half-depth $\langle d \rangle_{t}$ -- as defined in equation \eqref{eq:def_interface_thicknesses} -- represents an appropriate measure of the mixing zone as a region of strong interactions between the two fluid layers. 
We discuss further implications of these vertical profiles (including for values of the average flux coefficient) in more detail in section \ref{sec:Discussion}.}

\subsection{Convergence of vertical stirring, dissipation and mixing for extended domains}
\label{subsec:Convergence_of_vertical_stirring_and_dissipation_for_extended_domains}

%------------------------------------------------------------------------------------------
\begin{figure}
\centering
\includegraphics[scale = 1.0]{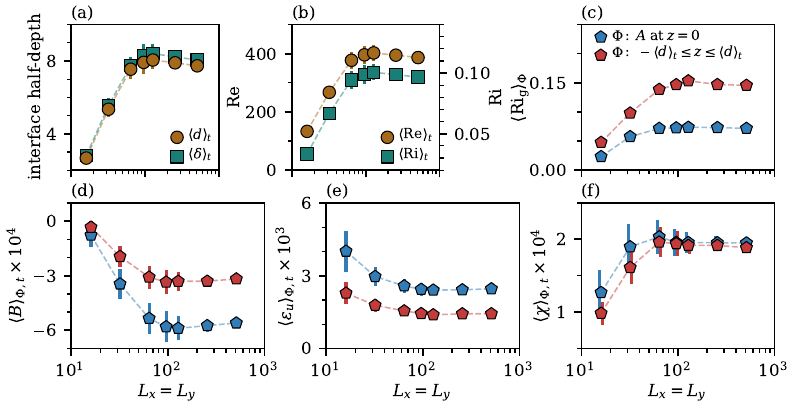}
\caption{\justifying{Impact of the horizontal extent of the domain on the mixing.
The flow topology 
-- comprising the 
(a) interface (half-) depths, 
(b) final emergent (bulk) Reynolds number and Richardson number, 
as well as averages across the midplane and across the entire mixing zone of 
(c) gradient Richardson number,
(d) buoyancy flux,
(e) kinetic energy dissipation and 
(f) buoyancy variance dissipation --
only converges for horizontally extended domains, $\Gh \gtrsim \Ghcrit = 96$.
Vertical solid lines indicate the temporal standard deviation.
\rev{All panels share the same abscissa.}
}}
\label{fig:impact_horizontal_extent_of_the_domain}
\end{figure}
%------------------------------------------------------------------------------------------

After having introduced important quantities related to the \rev{vertical interaction} of the velocity and buoyancy field in the previous sections \ref{subsec:Exemplary_evolution_of_a_forced_stratified_shear_flow} and \ref{subsec:Interaction_between_the_two_layers_of_fluid}, in figure \ref{fig:impact_horizontal_extent_of_the_domain} we plot their variation as a function of $\Gh$. 
Underlining the conclusions from figure \ref{fig:interaction_of_the_two_layers} (a), figure \ref{fig:impact_horizontal_extent_of_the_domain} (a) shows that the depths of the shear layer or the streamwise velocity and buoyancy interfaces indeed converge for \rev{sufficiently} extended horizontal domains. As the parameters $\left\{ \Re, \Ri \right\} \propto d$, they converge for extended domains, \rev{too,} as shown in panel (b). Note that both are significantly increased from their initial values. 
\rev{Starting with panel (c), we plot spatio-temporal averages of various quantities where the spatial averages are performed either across the two-dimensional horizontal midplane or across the volume of the entire turbulent mixing zone.}
Unsurprisingly, \rev{given the convergence of the depth of the mixing zone, we observe similar convergence for the gradient Richardson number $\Rig$} as well as the vertical stirring- and mixing-related quantities $B$, $\epsU$, and $\chi$ as shown in panels (c -- f). 

In summary, we find that the properties of the (predominantly vertical) stirring, dissipation and mixing actually depend strongly on the horizontal extent of the domain $\Gh$. However, the associated emergent vertical dynamics of the flow seems indeed to become independent of $\Gh$ once $\Gh \gtrsim \Ghcrit = 96$ where $\Ghcrit$ is a critical extent of the domain. 

%------------------------------------------------------------------------------------------
\begin{figure}
\centering
\includegraphics[scale = 1.0]{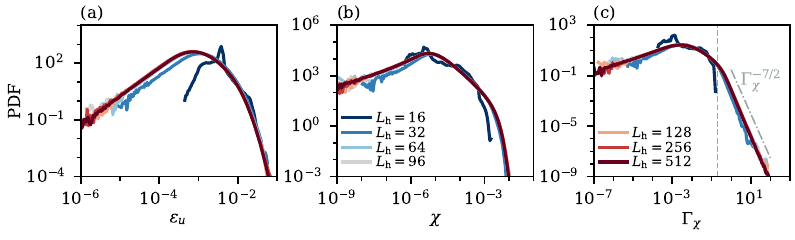}
\caption{\justifying{Statistics of mixing properties at midplane.
Statistical distributions of the 
(a) kinetic energy dissipation rate $\epsU$, 
(b) scaled buoyancy variance dissipation rate $\chi$, and 
(c) local flux coefficient $\Gchi$ 
are affected by insufficient extents of the domain \rev{but converge eventually}.
The grey dashed vertical line marks the canonical value $\Gchi = 0.2$. Note the emergent scaling for $\Gchi$ for extreme mixing events, and the marked difference of the high tails of the PDFs of $\epsU$ and $\chi$.
}}
\label{fig:energy_dissipation_at_midplane}
\end{figure}
%------------------------------------------------------------------------------------------

Finally, as shown in figure \ref{fig:energy_dissipation_at_midplane}, \rev{we remark that not only the mean values of} of dissipation and mixing rates \rev{depend} on $\Gh$, \rev{but also the general structure of their probability density functions (PDFs) depends on $\Gh$  as well}. 
Interestingly, we observe a pronounced scaling of the PDF of the \rev{(point-wise)} local flux coefficient $\Gchi$ -- defined as
\begin{equation}
\label{eq:def_GammaChi}
\Gchi := \frac{\chi}{\epsU} 
\end{equation}
and representing the local ratio between dissipation of scaled buoyancy variance and kinetic energy -- for values $\Gchi \gtrsim 0.2$, i.e., the canonical value proposed by \cite{osborn_1980}, as indicated in panel (c). 
\rev{We return to a more  conventional definition of a (in some sense) global flux coefficient -- based on average rather than local quantities -- in section \ref{sec:Discussion}.}

\subsection{Anisotropy of emergent large-scale dynamics}
\label{subsec:Anisotropy_of_emergent_large-scale_dynamics}

While the previous section \ref{subsec:Convergence_of_vertical_stirring_and_dissipation_for_extended_domains} has made clear that \textit{vertical} aspects of the flow dynamics converge for horizontally highly extended domains (only), section \ref{subsec:Exemplary_evolution_of_a_forced_stratified_shear_flow} has demonstrated that the considered forced stratified shear flows also develop certain characteristic \textit{horizontal} structures such as the primary Kelvin-Helmholtz billows. Therefore, it is appropriate to investigate whether there are at later times, when such early billows have broken down, emergent horizontally-aligned structures in the statistically steady flow as well. 

%------------------------------------------------------------------------------------------
\begin{figure}
\centering
\includegraphics[scale = 1.0]{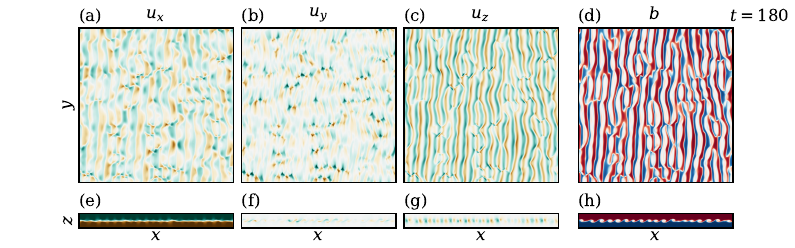}
\includegraphics[scale = 1.0]{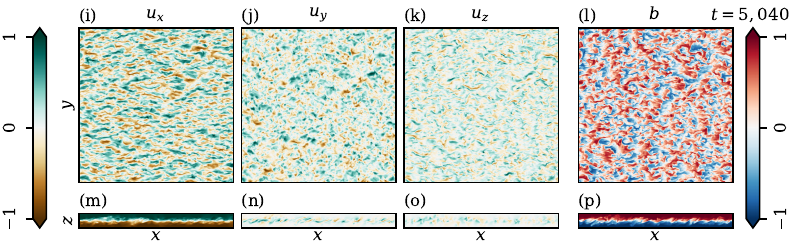}
\caption{\justifying{Emergent horizontally extended dynamics.
From (a -- h) early to (i -- p) late times, 
the size of emergent large-scale flow structures exhibits a strong anisotropy between their (a -- d, i -- l) horizontal and (e -- h, m -- p) vertical extension.
This figure shows data from the largest square domain, $\Gh = 512$, with $z=0$ in (a -- d, i -- l) or $y = 0$ in (e -- h, m -- p).
While the distinct streamwise \enquote{overturning billows} and spanwise \enquote{knots} and \enquote{tubes} mentioned in the introduction are prominent at early times, these structures (at least superficially) disappear at later times.
}}
\label{fig:emerging_horizontally_extended_dynamics}
\end{figure}
%------------------------------------------------------------------------------------------

Figure \ref{fig:emerging_horizontally_extended_dynamics} shows both the early and final emergent dynamics present in our most extended square domain of $\Gh = 512$. 
Panels (a -- d, i -- l) depict the entire horizontal cross-section at midplane, $\Phi \left( z = 0 \right)$, whereas panels (e -- h, m -- p) depict an associated vertical slice at $\Phi \left( y = 0 \right)$. 
Note that the vertical slices of $b$ from panels (h, p) \rev{are reminiscent of} figure \ref{fig:temporal_evolution} (c, e) despite the domain now being $16$ times as large. 
A video of the evolution of this flow -- from $t = 0$ to $t = \tevo$ -- is provided as supplemental material.

At the earlier time, as shown in panel (h), the flow is -- analogously to the simulation discussed in section \ref{subsec:Exemplary_evolution_of_a_forced_stratified_shear_flow} and shown in figure \ref{fig:temporal_evolution} -- clearly associated with the growth of the initial KHI and the subsequent \rev{non-linear} formation of overturning billows. 
Interestingly, as becomes clear via the associated horizontal slice in panel (d), although the initial overturning billows may extend across the entire extended spanwise direction, knots, tubes and defects between these rolls may introduce imperfections to this otherwise regular pattern. \rev{This effect has also} previously been observed experimentally \citep{thorpe_1985,thorpe_1987,caulfield_1996,thorpe_2002} and numerically \citep{fritts_2022a,fritts_2022b} for (in the spanwise direction) sufficiently wide domains. However, these early aspects of the flow dynamics appear (at least superficially) to be absent or smeared out by the sustained turbulence at late times as shown in panels (l, p).

As the velocity field, in particular $u_{x}$ as shown in panels (a, e, i, m), exhibits structures that are very similar to the buoyancy field $b$ shown in panels (d, h, l, p), the observations made in the previous sections are further supported. However, we find that both the initial pattern formation as well as the late sustained turbulence within the mixing zone can be recognised more easily from the scalar buoyancy field than from the vectorial velocity field.

Independently of the specific point in time, this comparison of vertical slices with the corresponding horizontal midplanes demonstrates the co-existence of  apparently quasi-regular dynamics in both the streamwise and spanwise directions. \rev{However, these  visualisations suggest that the emergent large-scale dynamics exhibits a strong scale separation and, thus, anisotropic associated length scales $\Lambda$. More specifically, at least qualitatively 
the horizontal structures seem to be much, much larger than the shear layer half-depth, i.e. $\left\{ d, \delta \right\} \ll \Lambda_{\textrm{h}}$.}

\begin{figure}
\centering
\includegraphics[scale = 1.0]{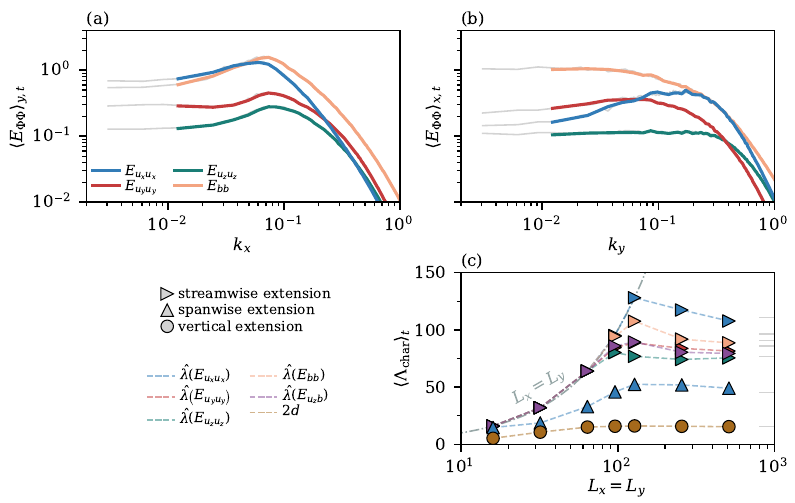}
\caption{\justifying{Magnitude quantification of emergent \rev{large-scale dynamics}.
(a) Emergent flow structures offer pronounced spectral peaks $\hat{\lambda}$ in the streamwise directions for all variables $\left\{ u_{x}, u_{y}, u_{z}, b\right\}$.
In contrast, (b) only the streamwise velocity $u_{x}$ exhibits a pronounced peak in spanwise direction.
(c) A systematic comparison of characteristic extensions of emergent flow structures along the streamwise, spanwise, and vertical directions highlights a strong anisotropy of the large-scale dynamics.
Note that while the coloured lines in panels (a, b) correspond to our largest square domain with $\Gh = 512$, grey lines are extracted from even more extended but non-square domains ($\Gx = 2048$ and $\Gy = 512$ or $\Gy = 2048$ and $\Gx = 512$).
Moreover, as the dash-dotted line $\Gx = \Gy$ in panel (c) demonstrates that flow structures may clearly be limited or affected by horizontally insufficiently extended domains, only horizontal extents $\Gh \gtrsim 256$ are large enough to resolve the  most extended flow structures in the horizontal direction.
}}
\label{fig:quantifying_the_size_of_emerging_flow_structures}
\end{figure}

As a first step towards a quantitative consideration  of the streamwise and spanwise extension of the emergent horizontal dynamics of the flow, we compute the Fourier (energy or co-) spectra 
\begin{alignat}{5}
E_{\Phi_{1} \Phi_{2}} \left( k_{x}, y, z = 0, t \right) &:= C \thickspace \real \left( \hat{\Phi}_{1}  \hat{\Phi}_{2}^{*} \right)
\qquad &&\textrm{with} \quad
\hat{\Phi}_{1, 2} \equiv \hat{\Phi}_{1, 2} \left( k_{x}, y, z = 0, t \right) , \\
E_{\Phi_{1} \Phi_{2}} \left( x, k_{y}, z = 0, t \right) &:= C \thickspace \real \left( \hat{\Phi}_{1}  \hat{\Phi}_{2}^{*} \right)
\qquad &&\textrm{with} \quad
\hat{\Phi}_{1, 2} \equiv \hat{\Phi}_{1, 2} \left( x, k_{y}, z = 0, t \right) 
\end{alignat}
of various quantities. Here, $\hat{\Phi}$ represent the Fourier coefficients corresponding to a transform in either the streamwise or spanwise direction, the asterisk $^{*}$ denotes the complex conjugate, and $k_{x}$ and $k_{y}$ are the associated streamwise and spanwise components of the wave vector, respectively. In order to allow for a direct comparability of these spectra with their corresponding terms in the kinetic energy or buoyancy variance equation 
via Parseval's theorem, the coefficient $C$ depends on $\Phi_{1}$ and $\Phi_{2}$ as follows: 
$C = 1/2$ for $\Phi_{1} = \Phi_{2} \in \left\{ u_{x}, u_{y}, u_{z}, b \right\}$, 
$C = \RiO$ for $\Phi_{1} = u_{z} \wedge \Phi_{2} = b$ (or vice versa), 
$C = 2 / \ReO$ for $\Phi_{1} = \Phi_{2} = \bm{S}$, or
$C = 1 / \left( \ReO \Pr \right)$ for $\Phi_{1} = \Phi_{2} = \nabla b$.
We summarise the key results  in figure \ref{fig:quantifying_the_size_of_emerging_flow_structures}.

%------------------------------------------------------------------------------------------
%------------------------------------------------------------------------------------------

In figure \ref{fig:quantifying_the_size_of_emerging_flow_structures} (a), we plot the streamwise energy spectra of $\left\{ u_{x}, u_{y}, u_{z}, b \right\}$, subject to appropriate spatio-temporal averages. 
Remarkably, we find that all these flow fields exhibit pronounced spectral peaks at streamwise wave numbers $\hat{k}_{x} \sim \mathcal{O} \left( 10^{-2} \right)$. This implies that a certain (narrow range of) streamwise wavelength(s) $\hat{\lambda}_{x} = 2 \pi / \hat{k}_{x}$ is particularly energetic. In other words, these pronounced spectral peaks establish the existence of flow structures with a characteristic streamwise length scale $\Lambda_{x} \sim \mathcal{O} \left( 10^{2} \right)$ in all of these flow fields, thus implying that there is a preferred length scale for the emergent streamwise self-organisation of the large-scale dynamics. 
We propose a potential dynamical origin of this characteristic scale $\Lambda_{x}$ in section \ref{sec:Discussion}.

As shown in figure \ref{fig:quantifying_the_size_of_emerging_flow_structures} (b), this emergent property of the streamwise spectra is in clear contrast to the behaviour of the spanwise spectra. 
On the one hand, we find a similarly pronounced spectral peak only for the $x$-component of the velocity field in the spanwise direction. Together with $\hat{k}_{y} \sim \mathcal{O} \left( 10^{-1} \right)$, this implies again the existence of a characteristic spanwise length scale $\Lambda_{y} \sim \mathcal{O} \left( 10^{1} \right)$.
On the other hand, the other flow fields do not exhibit such a peak but rather flatten out at small $k_{y}$, demonstrating that there is no preferred characteristic length scale for the spanwise self-organisation of the large-scale dynamics in $\left\{ u_{y}, u_{z}, b \right\}$.
\rev{We remark} that even though $E_{u_{y} u_{y}} ( k_{y, \textrm{min}} ) < E_{u_{y} u_{y}} ( \hat{k}_{y} )$, this difference is smaller than a factor of two and so we do not consider the associated maximum to be a \enquote{pronounced} spectral peak. 

It is natural to ask whether there could be additional spectral peaks at even smaller wave numbers, i.e., whether \rev{the large-scale dynamics might still be biased or limited by the finite domain of $\Gh = 512$.}
For this reason, we have conducted two additional simulations which increase either the streamwise or the spanwise extent of the domain by another factor of four. This results in $\Gx \times \Gy = 2048 \times 512$ or $512 \times 2048$, respectively. We include the associated energy spectra in figure \ref{fig:quantifying_the_size_of_emerging_flow_structures} (a, b) as grey solid lines. 
We find no evidence of such additional spectral peaks.
Furthermore, as both the location and amplitude of the peaks from these spectra coincide with the ones from our largest square domain, \rev{we are confident that there is strong evidence that these spectral peaks are  characteristic of this particular flow and, crucially, independent of the horizontal extent of the domain}. This implies that the large-scale dynamics is (at such large $\Gx$ and $\Gy$) indeed governed by mechanisms intrinsic to the flow. 
This is supported by appendix \ref{app:Convergence_of_horizontal_energy_spectra}, where we show that the energy spectra derived from smaller (yet sufficiently large) domains are also shown to converge with the present ones from the largest domains.

%---------------------------------------------------------------------------
\newcommand{\hp}{\hphantom{00}}
\newcommand{\hpp}{\hphantom{0}}
\newcommand{\gc}[1]{\textcolor{gray}{#1}}
\begin{table}
\centering
\begin{tabular}{@{\hskip 0mm} r @{\hskip 2.0mm} | c @{\hskip 3.5mm} c @{\hskip 3.5mm} c @{\hskip 3.5mm} c @{\hskip 3.5mm} c @{\hskip 2.0mm} | c @{\hskip 2.0mm} | c @{\hskip 3.5mm} c @{\hskip 0mm}}
\multicolumn{1}{c}{} & \multicolumn{5}{c}{streamwise $\hat{\lambda}$} & \multicolumn{1}{c}{spanwise $\hat{\lambda}$} & \multicolumn{2}{c}{vertical extent} \\
$\Gx \times \Gy \thickspace \thickspace$ & $E_{u_{x} u_{x}}$ & $E_{u_{y} u_{y}}$ & $E_{u_{z} u_{z}}$ & $E_{b b}$ & $E_{u_{z} b}$ & $E_{u_{x} u_{x}}$ & $2 d$ & $2 \delta$ \\
% This was the header - here comes the content.
  $16^{2}$ \hp     & \hpp \gc{$16 \pm \hpp 0$} & \gc{$16 \pm \hpp 0$} & \gc{$16 \pm \hpp 0$} & \hpp \gc{$16 \pm \hpp 0$} & \gc{$16 \pm \hpp 0$} & \gc{$15 \pm \hpp 3$} & \hpp \gc{$5.3 \pm 0.3$} & \hpp \gc{$5.6 \pm 0.3$} \\
  $32^{2}$ \hp     & \hpp \gc{$32 \pm \hpp 0$} & \gc{$32 \pm \hpp 0$} & \gc{$32 \pm \hpp 0$} & \hpp \gc{$32 \pm \hpp 0$} & \gc{$32 \pm \hpp 0$} & \gc{$19 \pm \hpp 6$} &     \gc{$10.7 \pm 0.7$} &     \gc{$11.1 \pm 0.7$} \\
  $64^{2}$ \hp     & \hpp \gc{$64 \pm \hpp 0$} & \gc{$64 \pm \hpp 0$} & \gc{$64 \pm \hpp 2$} & \hpp \gc{$64 \pm \hpp 0$} & \gc{$64 \pm \hpp 0$} & \gc{$33 \pm     14$} &         $15.1 \pm 0.9$  &     \gc{$15.5 \pm 0.7$} \\
  $96^{2}$ \hp     & \hpp \gc{$95 \pm \hpp 7$} &     $84 \pm     20$  &     $80 \pm     23$  & \hpp \gc{$94 \pm \hpp 9$} &     $86 \pm     20$  &     $46 \pm     24$  &         $15.8 \pm 1.1$  &         $16.5 \pm 1.1$  \\
 $128^{2}$ \hp     &     \gc{$128 \pm \hpp 0$} &     $89 \pm     32$  &     $77 \pm     27$  &         $108 \pm     30$  &     $89 \pm     32$  &     $52 \pm     24$  &         $16.1 \pm 0.8$  &         $16.8 \pm 0.9$  \\
 $256^{2}$ \hp     &         $117 \pm     18$  &     $84 \pm     13$  &     $74 \pm     12$  & \hpp     $92 \pm     20$  &     $80 \pm     12$  &     $52 \pm     27$  &         $15.8 \pm 0.4$  &         $16.4 \pm 0.3$  \\
 $512^{2}$ \hp     &         $108 \pm     19$  &     $82 \pm     10$  &     $76 \pm \hpp 9$  & \hpp     $89 \pm     11$  &     $79 \pm \hpp 8$  &     $49 \pm     15$  &         $15.5 \pm 0.2$  &         $16.1 \pm 0.1$  \\
\\
$2048 \times  512$ &         $111 \pm     17$  &     $86 \pm     12$  &     $77 \pm     10$  & \hpp     $96 \pm     12$  &     $91 \pm     10$  &     $46 \pm     10$  &         $15.6 \pm 0.1$  &         $16.2 \pm 0.1$  \\
 $512 \times 2048$ &         $118 \pm     13$  &     $83 \pm     10$  &     $75 \pm \hpp 8$  & \hpp     $95 \pm     11$  &     $86 \pm     12$  &     $45 \pm     15$  &         $15.5 \pm 0.1$  &         $16.2 \pm 0.1$
\end{tabular}
\caption{\justifying{Emergent dynamical \rev{length} scales.
This table quantifies the streamwise, spanwise, and vertical extent of the emergent large-scale dynamics via the wavelength associated with the spectral peaks and \rev{total depths of} the shear layer or density interface. 
We list values of the temporal mean and standard deviation, the latter of which might be significant due to the discrete nature of wave numbers. 
Unreliable values, i.e. those clearly affected by the (insufficiently extended) finite domain, are displayed in grey.
}}
\label{tab:emergent_dynamical_scales}
\end{table}
\let\hp\undefined
\let\hpp\undefined
\let\gc\undefined
%--------------------------------------------------------------------------

Quantifying characteristic length scales associated with the emergent large-scale dynamics, we extract the wavelengths corresponding to these pronounced spectral peaks (or these spectra's global maxima) for each simulation and summarise them in table \ref{tab:emergent_dynamical_scales}. 
Additionally, figure \ref{fig:quantifying_the_size_of_emerging_flow_structures} (c) shows this dependence on $\Gh$.
For small $\Gh$, as highlighted by the dash-dotted line, these extracted horizontal length scales $\Lambda_{\left\{ x, y \right\}}$ are clearly biased by the extent of the domain. However, this changes once $\Gh \gtrsim \Ghcrit$ with the critical horizontal extent $\Ghcrit$ depending on the solution field (and, thus, the associated final characteristic length scale). 
While most of the peaks are already properly represented given $\Ghcrit = 96$, some of them -- such as $\hat{\lambda}_{x} \left( E_{u_{x} u_{x}} \right)$ -- may require $\Gh = 256$ to be fully resolved. 
Hence, \rev{for the set of control parameters considered here,} a full convergence of the characteristic length scales associated with the large-scale dynamics is reached for  $\Gh \gtrsim 256$ only. 
\rev{This convergence of finite horizontal scales is, at least qualitatively, consistent with the convergence of the finite vertical depth of the shear layer as addressed in the previous sub-sections.}

In addition to the characteristic streamwise and spanwise length scales associated with the large-scale dynamics, we include their characteristic vertical extent via the total shear layer depth in figure \ref{fig:quantifying_the_size_of_emerging_flow_structures} (c). Note that since $d \approx \delta$ or $R \approx 1$, these data points equivalently show the total buoyancy interface depth. 
Comparing the characteristic length scales in the three spatial directions (as denoted by different symbol types), there is a clearly apparent anisotropy of the large-scale dynamics. 
While vertical scales $\Lambda_{z} \approx 16$ (circles) and spanwise scales $\Lambda_{y} \approx 50$ (upright triangles), streamwise scales (sideways triangles) may extend up to $\Lambda_{x} \approx 115$. Consequently, we find a hierarchy $\Lambda_{z} < \Lambda_{y} < \Lambda_{\textrm{x}}$ that spans across one order of magnitude.

Having extracted these characteristic scales of large-scale dynamics, there are two particularly striking observations. 
Firstly, while $u_{z}$ and $b$ are related to the vertical transport of buoyancy via equation \eqref{eq:def_vertical_buoyancy_advection}, their individual scales $\hat{\lambda}_{x}$ in $E_{u_{z} u_{z}}$ and $E_{b b}$ differ significantly. 
Secondly, the total depth of the mixing zone, representing the characteristic vertical extent $\Lambda_{z}$, actually converges at smaller domain extents $\Gh$ compared with $\Lambda_{x}$ from $E_{b b}$.
These two observations suggest the natural question as to why or how can the vertical characteristics (in both the velocity and buoyancy field) converge before the domain is sufficiently large to allow the separate flow fields to converge horizontally?

%------------------------------------------------------------------------------------------
\begin{figure}
\centering
\includegraphics[scale = 1.0]{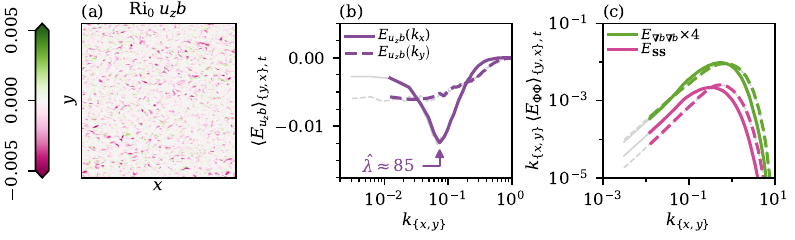}
\caption{\justifying{Buoyancy exchange and dissipation.
(a) The buoyancy flux (i.e. the vertical buoyancy advection) across the midplane 
--  $B \left( x, y, z = 0, t = \tevo \right)$ for $\Gh = 512$ -- 
generally stabilises the configuration. 
(b) The associated streamwise co-spectrum establishes the existence of a characteristic corresponding scale via a pronounced peak, which is in contrast to the spanwise direction. 
(c) Most of the dissipation, however, is associated with smaller scales similar to the mixing zone depth. 
In panels (b, c), coloured and grey spectra  correspond -- similar to figure \ref{fig:quantifying_the_size_of_emerging_flow_structures} (a, b) -- to the largest square and non-square domains, respectively.
}}
\label{fig:exchange_between_the_two_layers_of_fluid}
\end{figure}
%------------------------------------------------------------------------------------------

To answer this question, we visualise the buoyancy flux at the midplane in figure \ref{fig:exchange_between_the_two_layers_of_fluid} (a). Remembering that $B < 0$ when $u_{z}$ and $b$ have opposite signs -- i.e., dense parcels are moving upwards or light parcels are moving downwards --, consistently with figure \ref{fig:impact_horizontal_extent_of_the_domain} (d), it is reasonable to expect that the field shown in figure \ref{fig:exchange_between_the_two_layers_of_fluid} (a) is negative on average. 
Despite its complexity, this visualisation of $B$ suggests again the presence of some regularity.
While this is confirmed via the associated streamwise co-spectrum, as shown in panel (b), there is no such peak in the spanwise direction. 
We remark that a peak in this co-spectrum indicates the presence of a scale at which $u_{z}$ and $b$ interact most strongly, or, in other words, a scale at which vertical velocity and buoyancy are strongly correlated (or anti-correlated).
This interaction scale $\hat{\lambda}_{x} \left( E_{u_{z} b} \right) \approx 85$ with $\hat{\lambda}_{x} \left( E_{u_{z} u_{z}} \right) < \hat{\lambda}_{x} \left( E_{u_{z} b} \right) < \hat{\lambda}_{x} \left( E_{b b} \right)$. 
Hence, this peak (at least partially) explains the convergence of the vertical stirring and mixing for $\Gh \gtrsim \Ghcrit \gtrsim \hat{\lambda}_{x} \left( E_{u_{z} b} \right)$ despite other parts of the large-scale dynamics not yet being fully resolved.

%------------------------------------------------------------------------------------------
\section{Discussion}
\label{sec:Discussion}

\rev{As presented in section \ref{sec:Results} via a series of direct numerical simulations in domains of varying horizontal extent, the statistically stationary regime -- which can be entered by stably stratified shear flows after an initial transient if  subjected to a continuous forcing -- can indeed be associated with a strongly hierarchical anisotropy of the finite-size large-scale dynamics. 
As shown in section \ref{subsec:Convergence_of_vertical_stirring_and_dissipation_for_extended_domains}, a convergence of vertical stirring- and mixing-related properties can be reached once the vertical extent of the mixing zone, \rev{as introduced in section \ref{subsec:Interaction_between_the_two_layers_of_fluid}}, has converged. For the set of control parameters considered here, this has been the case for $\Gh \gtrsim \Ghcrit = 96$.} 
However, expecting dissipation to be dominated by the smallest scales in the flow, it might be surprising to see that the dissipation (and mixing) statistics only converge for $\Ghcrit \gg \etaK$ (with the Kolmogorov scale $\etaK \sim \mathcal{O} \left( 10^{-1} \right)$, see appendix \ref{app:resolving_shear_flows}). Questioning this potential expectation, we plot the energy spectra associated with the dissipation rates of kinetic energy and buoyancy variance in figure \ref{fig:exchange_between_the_two_layers_of_fluid} (c). Note that we show them in a pre-multiplied form \citep{Krug2020} in order to highlight visually those wave numbers that cause most of the variance. We find the peaks in these spectra to be located at $3 \times 10^{-1} \lesssim \hat{k} \lesssim 7 \times 10^{-1}$ or $9 \lesssim \hat{\lambda} \lesssim 18$, i.e., being again actually associated with the depth of the turbulent mixing zone. As we have shown in figure \ref{fig:exchange_between_the_two_layers_of_fluid} (b) and explained in section \ref{subsec:Anisotropy_of_emergent_large-scale_dynamics}, a convergence of this depth depends on resolving the streamwise spectral peak of the vertical buoyancy flux and requires $\Gh \gtrsim \Ghcrit$. Hence, the convergence of statistical aspects of dissipation goes hand in hand with the macroscopic or structural convergence of the turbulent mixing zone.

\rev{This convergence of the depth of the turbulent mixing zone to a finite characteristic value $\Lambda_{z} = 2 \langle d \rangle_{t} \gg 2$ (remember that we measure lengths in units of $\dO$) for sufficiently extended domains, despite the ongoing forcing and mixing, is a central result of section \ref{subsec:Interaction_between_the_two_layers_of_fluid}.
Before this convergence was reached, we observed a growth of its half-depth with an increasing extent of the domain.} 
\rev{This growth can be explained by an increased number of degrees of freedom of the underlying dynamical system, leading to a higher potential complexity,  and thus more vigorous turbulence that causes both a stronger macroscopic stirring as well as a stronger microscopic mixing. 
However, this trend must eventually be limited at some point as such} stirring increases the strength of the stratification.

Simply, the deepening of the mixing zone cannot continue without limit since turbulence cannot be sustained when the stratification becomes too strong. 
\rev{This strength of stratification can be} quantified by either the bulk Richardson number $\Ri$ or appropriate averages of the gradient Richardson number $\Rig$, see again figure \ref{fig:impact_horizontal_extent_of_the_domain} (c). 
The particular observation that the (mixing zone) average of $\Rig$ is approximately $0.15$ is \rev{not only supported by the Miles-Howard criterion but also} highly reminiscent of the results reported by \cite{portwood_2019}. They used a control strategy (effectively through modulating gravity) to identify equilibrium turbulent states in linearly stratified flows driven by constant vertical shear. 
Although the flow considered here is different, the key dynamics appears to be similar: the forced flow adjusts until the stratification is as strong as possible to still allow for vigorous turbulence which is able to stir and mix the buoyancy field.

\rev{Hence, we hypothesise that forced stratified shear flows \enquote{tune} themselves towards equilibrium states associated with typical values of $\Rig \lesssim 0.2$, eventually governing the resulting depth of the turbulent mixing zone. 
This \enquote{tuning} is highly reminiscent of the \enquote{kind of equilibrium} proposed by \cite{turner_1973}, and indeed has points of connection with the more recently-proposed arguments of \enquote{self-organised criticality} in stratified shear flows \citep{salehipour_2018,smyth_2019}. This tuning was also observed in wall-bounded stratified plane Couette flow by \cite{zhou_2017}, where analogously, an apparently maximum possible midplane gradient Richardson number for sustained turbulence $\Rig \simeq 0.2$ was identified. \cite{zhou_2017} characterised the closeness of this number to the marginal Miles-Howard criterion of $1/4$ as being potentially \enquote{fortuitous}, and argued that it may be more appropriate to interpret this value in terms of emergent turbulent balances, consistently with the deep physical insights of \cite{turner_1973}.} 

\rev{To complement this identification of a characteristic vertical scale associated with the large-scale dynamics, we have also identified characteristic streamwise scales $75 \lesssim \Lambda_{x} \lesssim 115$ associated with the large-scale dynamics.} We believe that this characteristic length scale can once again be interpreted in terms of the (sustained) emergent shear layer half-depth $\langle d \rangle_t \approx 8$. 
We remark that, although the imposed forcing is in principle designed to relax the flow back to the initial shear layer half-depth $\dO$ (and buoyancy interface half-depth $\deltaO$), the (above described) sustained turbulence  ensures this long-time significantly deeper shear layer and buoyancy interface. 
It is well-known (see for example \cite{scinocca_2000}) that the most unstable mode of KHI has a characteristic wavelength $\lambda_{\textrm{KHI}}$ of approximately fifteen times the shear layer half-depth. Given $\langle d \rangle_t \approx 8$ at late times, this implies $\lambda_{\textrm{KHI}} \approx 120$ which is very similar to the emergent streamwise length scales, particularly those associated with the streamwise velocity field as listed in table \ref{tab:emergent_dynamical_scales}.

\rev{This observation encourages us to conduct a classical (unforced) linear stability analysis -- following the procedures outlined in \citep{Smyth11, Smith2021} -- of the emergent background profiles $\langle u_{x} \rangle_{A, t}$ and $\langle b \rangle_{A, t}$ together with the mean associated turbulent diffusivities of momentum and buoyancy as built on the approach of \cite{Kaminski19}. The required decomposition of the background flow from the turbulent fluctuations is discussed in more detail below. Indeed, it turns out that the mode of fastest (real) growth rate is associated with a wavelength $109 \lesssim \lambda_{\textrm{background KHI}} \lesssim 121$. This highlights that even a sustained turbulent interface can be susceptible to a Kelvin-Helmholtz instability, with crucial implications for the emergent large-scale dynamics.}

However, this preferred instability scale is not able to roll up completely into coherent Kelvin-Helmholtz billows, and in particular, the turbulence certainly disrupts any possibility of subharmonic pairing or merging occuring, perhaps analogously to the disruption arguments put forward by \cite{mashayek_2013}. Furthermore, our observation that spanwise scales are both significantly smaller and significantly harder to identify is consistent with the complete lack of billow coherence, as the knot/tube/defect structure is observed experimentally to require billows of at least $3-5 \lambda_{\textrm{KHI}} \sim 360-600$ spanwise extent, which is an order of magnitude larger than the observed spanwise scale $\Lambda_{y} \approx 50$.
\rev{We remark that, both due to their temporal stationarity and absolute size, the here extracted large-scale dynamics appears to differ significantly and qualitatively from the inherently transient structures that monotonically \enquote{increase with time approximately from $5 \dO$ to $9 \dO$} in an unforced configuration as described in \citep{Watanabe2021}.}

%------------------------------------------------------------------------------------------
\begin{figure}
\centering
\includegraphics[scale = 1.0]{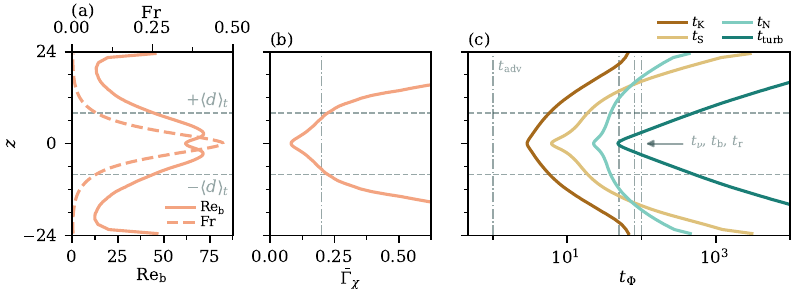}
\caption{\justifying{Additional characteristics of the emergent dynamics. 
The mixing zone is characterised by: (a) heightened values of $\Reb \gtrsim 35$ and $\Fr$;  
(b) values of the  bulk mixing coefficient smaller than the canonical (bounding) value of $\GchiBar = 0.2$ (marked by the vertical line).
(c) A comparison of present dynamically manifesting and global time scales highlights that the vast scale separation in space is complemented by another one in time as well.
Here, the flow has $\Gh = 128$, similarly to figure \ref{fig:interaction_of_the_two_layers}.
}}
\label{fig:Characteristics_of_emergent_dynamics}
\end{figure}
%------------------------------------------------------------------------------------------

\rev{The sustained large-scale dynamics is associated with a -- relative to its initial value or profile -- significantly increased gradient Richardson number $\Rig$ as shown in figure \ref{fig:interaction_of_the_two_layers} (b). However, this single measure is (of course) insufficient to characterise stratified shear flows. Two other commonly used parameters quantifying the relative strengths of turbulence, stratification and viscosity are the buoyancy Reynolds number
\begin{align}
\label{eq:def_buoyancy_Reynolds_number}
\Reb \left( z \right) := \ReO \frac{\langle \epsU \rangle_{A, t}}{\langle N^{2} \rangle_{A, t}} \equiv \left( \frac{\tN}{\tK} \right)^{2} 
\end{align}
and the (turbulent) %\pv{(I would ask to delete this "turbulent" since we do not consider the turbulent fields here.)} 
Froude number
\begin{align}
\label{eq:def_Froude_number}
\Fr \left( z \right) := \frac{\langle \epsU \rangle_{A, t}}{\langle E_{\textrm{kin}} \rangle_{A, t} \langle N \rangle_{A, t}} \equiv \frac{\tN}{\tturb}
\end{align}
where $E_{\textrm{kin}} := \left( u_{x}^{2} + u_{y}^{2} + u_{z}^{2} \right) /2$ is the (pointwise and total) kinetic energy (density). 
Figure \ref{fig:Characteristics_of_emergent_dynamics} (a) shows the vertical profiles of these two parameters $\Reb$ and $\Fr$ for the same flow as shown for $\Rig$ in figure \ref{fig:interaction_of_the_two_layers} (b).}

\rev{Interestingly, this figure suggests an (effectively equivalent) alternative interpretation of the mixing zone as the region where the buoyancy Reynolds number has a heightened value of $\Reb \gtrsim 35$.  
We remark that $\Reb$ can also be written as a ratio of length scales $\Reb \equiv \left( \etaO / \eta_{\textrm{K}} \right)^{4/3}$ where $\eta_{\textrm{K}}$ is the Kolmogorov microscale and $\etaO := \sqrt{\langle \epsU \rangle_{A, t} / \langle N^{3} \rangle_{A, t} }$ is the Ozmidov length scale. 
Here $\eta_{\textrm{K}}$ is assumed to vary with height (as introduced and plotted in appendix \ref{app:resolving_shear_flows}) and thus calculated based on $\langle \epsU \rangle_{A, t}$.
Since $\etaO$ can be interpreted as the characteristic scale of the largest eddy not significantly affected by the stratification whereas $\eta_{\textrm{K}}$ is the scale at which viscous dissipation completely damps motions, such larger values of $\Reb$ give at least some opportunity of a range of turbulent scales neither strongly affected by stratification nor by viscosity and thus allow (potentially) for vigorous turbulence and mixing. 
Dating back to the seminal work of Gibson (see for example \cite{gibson1987fossil}), there is much evidence that such values of $\Reb$ are associated with a recognisably turbulent stratified mixing dynamics, see for example \cite{Smyth2000} for a further discussion. Indeed, \cite{Shih05} argued that $7 \lesssim \Re \lesssim 100$ defined a transitional mixing regime with essentially constant bulk flux coefficient close to the canonical value of 0.2.}

\rev{We also see that the mixing zone is associated with heightened values of $\Fr$, perhaps unsurprising after consideration of figure \ref{fig:interaction_of_the_two_layers} (b) since there is a commonly assumed correlation between $\Rig \propto \Fr^{-2}$. However, caution should be exercised when considering particular numerical values of $\Fr$ as $\Rig \lesssim 0.25$ in the mixing zone, and so it is appropriate to think of the (shear-driven) mixing to be relatively weakly stratified and \enquote{dominated} by the shear \citep{Mater2014}.}

\rev{Furthermore, as we find an average of $\etaO \approx 6$ across the turbulent mixing zone, we recover the (length scale) hierarchy $\etaK \ll \etaO \ll \Lambda_{z} \ll \Lambda_{y} \ll \Lambda_{x}$, entirely consistently with the observation that there are no large-scale overturnings of the entire mixing zone. Although this observed hierarchy of scales in this self-organising emergent flow is consistent with that observed in the strongly stratified regime (sometimes also called the layered anisotropic stratified turbulence or \enquote{LAST} regime) considered by \cite{brethouwer2007scaling}, the dominant cause for anisotropy in this flow is the continually imposed shear forcing which leads (at least within the mixing zone) to relatively weakly stratified shear-driven (and shear-dominated) turbulence.}

\rev{Quantitative evidence of this relatively weakly stratified shear-driven turbulence can be obtained from consideration of mixing quantities. In section \ref{subsec:Convergence_of_vertical_stirring_and_dissipation_for_extended_domains} we introduced the (point-wise determined) local flux coefficient $\Gchi$ to highlight that even statistical aspects of the flow are affected by the large-scale dynamics. However, it is more common to describe this ratio of dissipation rates via average values of $\epsU$ and $\chi$.} Hence, we additionally define a bulk mixing coefficient
\begin{equation}
\label{eq:def_GammaBar}
\GchiBar := \frac{ \langle \chi \rangle_{A, t}}{ \langle \epsU \rangle_{A, t}} 
\end{equation}
and plot its vertical profile in figure \ref{fig:Characteristics_of_emergent_dynamics} (b). Note the relation of this panel to figure \ref{fig:interaction_of_the_two_layers} (d). Using the approach described in section \ref{subsec:Convergence_of_vertical_stirring_and_dissipation_for_extended_domains}, we find $\GchiBar \left( z = 0 \right) \simeq 0.08$ at the midplane and $\langle \GchiBar \rangle_{- \langle d \rangle_{t} \leq z \leq \langle d \rangle_{t}} \simeq 0.13$ within the turbulent mixing zone, essentially because $\epsU$ is more localised at the midplane than $\chi$ is. 
These values are somewhat smaller than the canonical value of $0.2$ proposed by \citet{osborn_1980}, \rev{though it is important to remember that this canonical value is a maximum rather than the more commonly assumed constant value. Within Osborn's arguments this maximum value of $\Gamma \leq 0.2$ is  associated with typical empirically observed maximum values of the \enquote{flux} Richardson number $\Rif \simeq \GchiBar/(1+\GchiBar)$, implying $\Rif \lesssim 0.17$ within our mixing zone. As demonstrated for example by \cite{zhou_2017} -- see also more detailed and pedagogical discussions in the review of \cite{caulfield_2021} --, $\Rif \simeq \Rig$ for weakly stratified shear-driven mixing and so this maximum value can be thought of as an appropriate maximum value for the \enquote{gradient} Richardson number as well.} 
\rev{As shown in figure \ref{fig:impact_horizontal_extent_of_the_domain} (c), $\Rig \simeq 0.15$ when averaged across the mixing zone and so it would seem superficially reasonable that $\GchiBar$ should be larger, since if $\Rif \simeq \Rig$ then $\GchiBar \simeq \Rig/(1-\Rig) \simeq 0.18$. }

\rev{However, there are at least two plausible reasons for the discrepancy. 
First, and perhaps most importantly, central to Osborn's model is the assumption that mixing (however quantified) is linearly proportional to the dissipation of turbulent kinetic energy. In our flows, it is clear that although both $\epsU$ and $\chi$ are elevated in the mixing zone, the assumption that they are linearly related is clearly neither qualitatively nor quantitatively correct.
The second reason is somewhat more subtle and related to the specific properties of the flow forcing being used by us here. 
As $\chi$ actually involves diffusion explicitly, it seems most natural to define a flux coefficient in terms of $\chi$ (and potentially its averages) rather than in terms of the buoyancy flux $-B$ which is not even sign-definite, although \cite{osborn_1980} did define the flux coefficient using $-B$ in the numerator. 
The two quantities $-B$ and $\chi$ balance each other in an appropriately constructed buoyancy variance equation when there is no forcing as discussed, for example, in \cite{caulfield_2021}, making the choice of $\chi$ or $-B$ in the numerator of $\Gamma$ moot. However, this balance does not occur in the presence of forcing as is apparent from the data shown in figures \ref{fig:interaction_of_the_two_layers} (c, d). 
For the present flow, we find $- \langle B \rangle_{A, t} / \langle \chi \rangle_{A, t} \approx 3$ at midplane or $\approx 1.5$ when averaged across the turbulent mixing zone. 
This mismatch is entirely due to the particular form of the forcing which we are using. The forcing of both the velocity and buoyancy fields to relax back to their original profiles inevitably perturbs in nontrivial ways both the kinetic energy and the potential energy, as well as (crucially) the direct exchange between these two reservoirs, and hence the buoyancy flux. This (admittedly somewhat artificial) observation underlines that, for our forced configuration, $\chi$ -- which naturally continues to describe the irreversible mixing processes -- is the appropriate reference for the definition of a flux coefficient $\Gamma$, although this choice may contribute to the numerical discrepancy between our observed $\GchiBar$ and expected value within the theoretical framework of \cite{osborn_1980}. }

\rev{Throughout this work, when considering the flow dynamics, we have always used the total solution fields. 
(Furthermore, we have defined buoyancy in terms of density deviations from the constant reference density rather than relative to the statically stable density field.) 
An alternative perspective is to decompose the fields into some background (or appropriately averaged) flow and the residual turbulent fluctuations, i.e., $\Phi = \bar{\Phi} + \Phi^{'}$ where the overbar denotes an average over which the fluctuations vanish. For our configuration, the appropriate decomposition is 
\begin{alignat}{5}
\label{eq:def_decomposition_background_vs_fluctuations_ux}
u_{x}   &= \bar{u}_{x}  &&+  &&u_{x}^{'} &
\quad &\textrm{with} &\thickspace \thickspace \bar{u}_{x} \left( z \right) &:= \langle u_{x} \rangle_{A, t}
, \\
u_{y}   &=              &&   &&u_{y}^{'}, \\
u_{z}   &=              &&   &&u_{z}^{'}, \\
\label{eq:def_decomposition_background_vs_fluctuations_b}
b       &= \bar{b}      &&+  &&b^{'} &
\quad &\textrm{with} &\thickspace \thickspace \bar{b} \left( z \right) &:= \langle b \rangle_{A, t}
\end{alignat}
where the time-independent background profiles emerge dynamically due to the continuous forcing defined in equations \eqref{eq:def_fu} and \eqref{eq:def_fb}. Considering the  dissipation rates as defined in equations \eqref{eq:def_dissipation_rate_kinetic_energy} and \eqref{eq:def_dissipation_rate_buoyancy_variance}, this leads to 
\begin{equation}
\epsU = \bar{\varepsilon}_{u} + \epsU^{'} + 2 \epsU^{\times}
\quad \textrm{with} \quad
\begin{cases}
\bar{\varepsilon}_{u} := \frac{2}{\ReO} \bar{\bm{S}}^{2} 
&\equiv \frac{1}{\ReO} \left( \frac{\partial \bar{u}_{x}}{\partial z} \right)^{2}\\
\epsU^{'} := \frac{2}{\ReO} \bm{S}^{' 2} \\
\epsU^{\times} := \frac{2}{\ReO} \bar{\bm{S}} \bm{S}^{'} 
&\equiv \frac{1}{\ReO} \frac{\partial \bar{u}_{x}}{\partial z} \left( \frac{\partial u_{x}^{'}}{\partial z} + \frac{\partial u_{z}^{'}}{\partial x} \right) , \\
\end{cases}
\end{equation}
\begin{equation}
\epsB = \bar{\varepsilon}_{b} + \epsB^{'} + 2 \epsB^{\times}
\quad \textrm{with} \quad
\begin{cases}
\bar{\varepsilon}_{b} := \frac{1}{\ReO \Pr} \left( \nabla \bar{b} \right)^{2} 
&\equiv \frac{1}{\ReO \Pr} \left( \frac{\partial \bar{b}}{\partial z} \right)^{2} \\
\epsB^{'} := \frac{1}{\ReO \Pr} \left( \nabla b^{'} \right)^{2} \\
\epsB^{\times} := \frac{1}{\ReO \Pr} \left( \nabla \bar{b} \right) \left( \nabla b^{'} \right) 
&\equiv \frac{1}{\ReO \Pr} \frac{\partial \bar{b}}{\partial z} \frac{\partial b^{'}}{\partial z} \\
\end{cases}
\end{equation}
with the strain rate tensor $\bm{S} = \bar{\bm{S}} + \bm{S}^{'}$.
By definition, the mixed contributions vanish on average, i.e. $\langle \epsU^{\times} \rangle_{A, t} = 0$ and $\langle \epsB^{\times} \rangle_{A, t} = 0$. In contrast, the vast majority  of the dissipation is associated with the turbulent fluctuations. We find that $\langle \epsU^{'} \rangle_{A, t} / \langle \epsU \rangle_{A, t} \approx \langle \epsB^{'} \rangle_{A, t} / \langle \epsB \rangle_{A, t} \gtrsim 0.8$ at all heights and the minimum happening at the midplane, perhaps unsurprisingly as that is the location of the largest vertical gradient in the background fields. 
Hence, these observations give us confidence that our calculations based on the total fields (incorporated, e.g., in $\Rig$, $\Reb$, and $\Fr$) enable equivalent insights for the turbulent fields.}

\rev{Finally, in order to highlight the (sub-)dominance of various mechanisms in our forced stratified shear flow, we compare various time scales in figure \ref{fig:Characteristics_of_emergent_dynamics} (c). 
Note that, based on our non-dimensionalisation from equation \eqref{eq:def_non-dimensionalisation}, all our non-dimensional time scales are related via $\tau_{\Phi} = t_{\Phi} \tauadv$ to their dimensional counterpart and, thus, scaled relative to the dimensional advective time scale $\tauadv$ of the background shear flow.
We consider the Kolmogorov, shear, buoyancy frequency, and turbulence decay time scales -- defined by 
\begin{equation}
\label{eq:def_time_scales_emergent}
\tK := \frac{1}{\sqrt{\ReO \langle \epsU \rangle_{A, t}}} , \quad
\tS := \frac{1}{\langle S \rangle_{A, t}} , \quad
\tN := \frac{1}{\langle N \rangle_{A, t}} , \thickspace \textrm{and} \quad
\tturb := \frac{\langle E_{\textrm{kin}} \rangle_{A, t}}{\langle \epsU \rangle_{A, t}} ,
\end{equation}
respectively -- as those time scales that are associated with the emergent flow.
The emergent non-dimensional (and vertically varying) parameters $\Rig$, $\Reb$, and $\Fr$ can then be interpreted as ratios of these timescales as shown on the right of equations \eqref{eq:def_gradient_Richardson_number}, \eqref{eq:def_buoyancy_Reynolds_number}, and \eqref{eq:def_Froude_number}.
Conversely, the advective, viscous diffusion, (bulk) buoyancy, and relaxation time scales -- defined by
\begin{equation}
\label{eq:def_time_scales_global}
\tadv := 1 , \quad
\tnu := \ReO , \quad 
\tb := \frac{1}{\RiO},  \thickspace \textrm{and} \quad
\tr ,
\end{equation}
respectively -- represent \enquote{global} time scales which are determined by the externally imposed control parameters. 
In general, the Batchelor and the buoyancy diffusion time scale, $\tB := \tK / \sqrt{\Pr}$ and $\tkappa := \ReO \Pr$, would be two further important time scales. However, since $\Pr \equiv \tkappa / \tnu = \left( \tK / \tB \right)^{2} = 1$, $\tB = \tK$ as well as $\tkappa = \tnu$ here. 
Figure \ref{fig:Characteristics_of_emergent_dynamics} (c) highlights a hierarchy  $\tadv < \tK < \tS < \tN < \tnu < \tb < \tr$ across the entire turbulent mixing zone while $\tturb \approx \tnu$ at midplane but $\tr \ll \tturb$ with increasing distance. 
As processes associated with shorter time scales tend to be stronger, this underlines that any effects introduced by the volumetric forcing can be considered sub-dominant. Moreover, finding $\tS < \left\{ \tN , \tturb \right\}$ underlines yet again that our forced flow lies in a weakly stratified shear-dominated regime \citep{Mater2014, caulfield_2021}.}

%------------------------------------------------------------------------------------------
\section{Conclusions}
\label{sec:Conclusions}

\rev{Motivated by their relevance to many geophysical flows, we have conducted direct numerical simulations of forced stratified shear flows to improve our fundamental understanding of their dynamics.}
Although our flow configuration is prone to a primary Kelvin-Helmholtz instability, it additionally and importantly includes a continuous forcing which tends to restore the initial profiles and ensures that the ensuing turbulence can be sustained for arbitrarily long times. Since it is this continuous forcing which connects to the motivating geophysical flows, we largely disregard the initial transient but rather focus on the \rev{regime of} statistically stationary dynamics during later times.

\rev{One of our objectives was to determine whether or not the resulting shear layer depth $d$ converges to a finite value -- which is independent of the extent of the domain and thus characteristic of the emergent self-organisation of the flow -- despite the ongoing forcing and mixing. To this end, in section \ref{sec:Results}, we have simulated this regime in numerical domains of varying horizontal extent in the range $16 \leq \Gh \leq 512$ (measured in units of the initial shear layer half-depth $\dO$). We found that the resulting total shear layer depth indeed converges to a finite value $\Lambda_{z} = 2 \langle d \rangle_{t} \approx 16$ once $\Gh \gtrsim 96$. Moreover, we have  also identified characteristic streamwise and spanwise length scales -- $75 \lesssim \Lambda_{x} \lesssim 115$ and $\Lambda_{y} \approx 50$, respectively -- associated with the emergent large-scale dynamics. As discussed in more detail in section \ref{sec:Discussion}, the emergent dynamics of this sustained turbulent flow can be associated with a  hierarchy of length and time scales -- $\etaK \ll \etaO \ll \Lambda_{z} \ll \Lambda_{y} \ll \Lambda_{x}$ and $\tadv < \tK < \tS < \tN < \tnu \lesssim \tturb < \tb < \tr$, respectively --, underlining that our self-organising emergent flow demonstrates strongly anisotropic behaviour.}

In summary, our research leads to two main conclusions. 
Firstly, \rev{as forced stratified shear flows result in a significantly enlarged yet finite shear layer depth,} it adds to an increasing body of evidence that such flows \enquote{tune} (in particular in the vertical direction) to a quasi-equilibirum state with typical values of $\Rig \lesssim 0.2$. Such a state appears to allow both sustained turbulence and non-trivial buoyancy flux and attendant irreversible mixing in a fundamentally \enquote{weakly stratified} (and hence, due to the forcing, shear-dominated) regime. 
Secondly, \rev{as the most unstable mode of the emergent background flow has a length scale  that is very similar to the emergent length scale of the characteristic large-scale dynamics of the streamwise velocity field,} 
we conjecture that even such strongly turbulent flows as the flows consider here  may  retain an imprint of the (inherently linear) shear-driven primary instability scale.
\rev{From a more practical perspective,} these two conclusions imply that domains should be scaled using the finite \enquote{tuned} emergent vertical depth of the turbulent shear layer, which (during this tuning process) may be significantly larger than its initial value in order to yield converged and reliable statistics. \rev{If the large-scale pattern formation is of interest, the streamwise and spanwise extents may need to be even larger, requiring highly extended domains.}

Clearly, these conclusions need to be tested for other choices of the key parameters. It would be very instructive to investigate the sensitivity of the \enquote{tuning} process to the choice of the initial bulk Richardson number and Reynolds number. Moreover, an application to the ocean would clearly require the investigation of the sensitivity of the flow dynamics to the choice of $\Pr$, as thermally stratified water has $\Pr \sim \mathcal{O} \left( 10 \right)$. As we have seen, the emergent scales are both clearly related to the turbulent processes but also enormously larger than the smallest scales of that turbulence. A particular issue of interest is to consider larger initial $\ReO$ that would be unstable right from the beginning of the simulation, avoiding the (potentially confusing) phenomenon of the shear layer effectively doubling in depth \rev{due to molecular diffusion} before the onset of the primary instability. If our conclusions prove to be robust and generic for other parameter choices, they have major implications for idealized computational studies, as they strongly suggest that to capture key characteristics of sheared turbulence in such relatively \enquote{weak} stratification, significantly extended computational domains are required. \rev{Furthermore, we have focussed exclusively on the end-member case of a continually forced shear layer, in contrast to the more widely-studied other end-member case of an unforced shear layer undergoing a \enquote{run-down} mixing event. Real flows will in general exhibit a transient forcing that lies somewhere between these two end members, and it is an interesting and open question as to how strongly dependent the flow dynamics would be on a finite time scale of forcing.}
 
There are also interesting implications for the geophysical application of such idealized studies of shear-driven stratified turbulence which conventionally completely ignore any effects of rotation. As our study suggests that streamwise scales can emerge that are $\mathcal{O} \left( 50 \right)$ times larger than the half-depth of the shear layer during the initial onset of KHI, it is entirely plausible that such larger scales will at least be affected somewhat by rotational effects. As noted in the review of \cite{taylor_2023}, \enquote{submesoscale motions} with Rossby number $\Ro := U/(fL) \sim O(1)$, where $f$ is the Coriolis parameter and $U$ and $L$ are characteristic velocity and length scales, are flows where \enquote{the Coriolis acceleration is important, but it does not constrain the motion}. In the world's oceans, they \enquote{define submesoscales as dynamical features with horizontal scales between approximately 200 m and 20 km}. It is entirely plausible that geophysical flows such as estuarine outflows could have initial shear layer half-depths of the order of 4-10 m, and that KHI would onset immediately associated with that shear layer half-depth as observed for example by \cite{holleman_2016}. In such flows, the emergent streamwise length scales may well experience non-trivial rotational effects. Therefore, it seems an interesting important question to investigate to what extent the emergent flow properties we have identified might be affected by larger-scale rotation.

%------------------------------------------------------------------------------------------
\backsection[Acknowledgements]{
P.P.V. 
(i) thanks Roshan J. Samuel (TU Ilmenau, Germany) for early help with NekRS and
(ii) gratefully acknowledges the support of Pembroke College, Cambridge, through a postdoctoral research associateship. 
The constructive and insightful comments of three anonymous referees have led to significant improvements in this document, and their helpful and thoughtful input is gratefully acknowledged.
}

\backsection[Funding]
{
P.P.V. is funded by the Deutsche Forschungsgemeinschaft (DFG, German Research Foundation) within Walter Benjamin Programme 532721742. 
The authors gratefully acknowledge the Gauss Centre for Supercomuting e.V. (\href{www.gauss-centre.eu}{\texttt{www.gauss-centre.eu}}) for funding this work by providing computing resources through the John von Neumann Institute for Computing (NIC) on the GCS supercomputer JUWELS at Jülich Supercomputing Center (JSC) within project nonbou.
}

\backsection[Declaration of interests]{
The authors report no conflict of interest.
}

\backsection[Author ORCIDs]{
P. P. Vieweg, https://orcid.org/0000-0001-7628-9902 \\
C. P. Caulfield, https://orcid.org/0000-0002-3170-9480 \\
}

\backsection[Author contributions]{
Both authors designed the study. P.P.V. performed the numerical simulations, processed the generated data, and drafted the manuscript. Both authors contributed equally in discussing the data and finalising the paper.
}

%------------------------------------------------------------------------------------------
\appendix

%------------------------------------------------------------------------------------------
\section{Resolving shear flows efficiently using spectral element methods}
\label{app:resolving_shear_flows}

Direct numerical simulations aim to resolve all dynamically relevant scales, from the domain size down to the Kolmogorov scale  or Batchelor scale. In our non-dimensional description, see again section \ref{subsec:Governing_equations} and equation \eqref{eq:def_non-dimensionalisation}, these smallest scales are locally given by
\begin{equation}
\label{eq:def_Kolmogorov_Batchelor_scale}
\eta_{\textrm{K}} := \left( \frac{1}{\ReO^{3} \thickspace \epsU} \right)^{1/4} 
\qquad \textrm{and} \qquad
\eta_{\textrm{B}} := \frac{\eta_{\textrm{K}}}{\sqrt{\Pr}} , 
\end{equation}
and both depend on the kinetic energy dissipation rate $\epsU$ as defined by equation \eqref{eq:def_dissipation_rate_kinetic_energy}.

%------------------------------------------------------------------------------------------
\begin{figure}
\centering
\includegraphics[scale = 1.0]{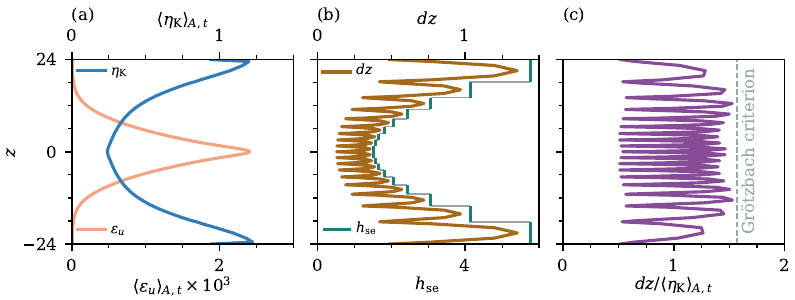}
\caption{\justifying{Resolving shear flows.
(a) Shear flows exhibit highly non-uniform profiles of dissipation $\epsU$ and, thus, the smallest dynamical scales $\etaK$.
(b) Spectral element methods allow for the adjustment of the height of spectral elements $h_{\mathrm{se}}$. The local resolution $dz$ follows from a subsequent spectral expansion of polynomial order $N$ within each element.
Hence, adjusting $h_{\mathrm{se}}$ enables (c) the very efficient resolution of shear flows.
Here, just like in figure \ref{fig:temporal_evolution}, $\Gh = 128$.
}}
\label{fig:resolving_shear_flows}
\end{figure}
%------------------------------------------------------------------------------------------

As already presented in figure \ref{fig:interaction_of_the_two_layers}, shear flows -- as visualised also in figure \ref{fig:temporal_evolution} -- offer highly non-uniform spatial distributions of dissipation, including this kinetic energy dissipation rate. 
For example, figure \ref{fig:resolving_shear_flows}  re-plots this $\langle \epsU \rangle_{A, t} \left( z \right)$ in panel (a). The dissipation is strongest at midplane but vanishes sufficiently far away. We compute the associated local Kolmogorov scale $\langle \eta_{\textrm{K}} \rangle_{A, t}$ from $\langle \epsU \rangle_{A, t}$ to obtain a proxy for the (on average) smallest local dynamical scale. 

The local (vertical) resolution $dz$ of spectral element methods is determined by the interplay between the size (here: height) of each spectral element $h_{\textrm{se}}$ and the polynomial order $N$ \citep{Deville2002, Vieweg2023a}.
It has empirically been shown \citep{Scheel2013} that spectral element methods offer smooth dissipation fields, even at the spectral element boundaries, when the (refined) Grötzbach criterion \citep{Scheel2013, Vieweg2023a}
\begin{equation}
\label{eq:Groetzbach_criterion}
\frac{dz}{\langle \eta_{\textrm{K}} \rangle_{A, t}} \lesssim \frac{\pi}{2} 
\quad \textrm{for } \thickspace \Pr \leq 1
\qquad \textrm{or} \qquad
\frac{dz}{\langle \eta_{\textrm{B}} \rangle_{A, t}} \lesssim \frac{\pi}{2} 
\quad \textrm{for } \thickspace \Pr \geq 1
\end{equation}
is satisfied. It is common to use this as a criterion for the spatial resolution.

Hence, as shown in panel (b), we adjust $h_{\textrm{se}} \left( z \right)$ to mimic the observed trends in $\langle \eta_{\textrm{K}} \rangle_{A, t} \left( z \right)$. After the spectral expansion within each spectral element, the local $dz$ follows a similar trend. 
As highlighted eventually by panel (c), this procedure of adjusting  the local spectral element height enables the successful and efficient resolution of shear flows.

Finally, we remark that we derived $h_{\textrm{se}} \left( z \right)$ from preliminary tests and use the same distribution for all our production simulations. The horizontal width of spectral elements $w_{\textrm{se}}$ is constant and similar to $h_{\textrm{se}} \left( z = 0 \right)$.
A comparison of time-averaged vertical profiles between our $\Gz = 48$ domain and a vertically even more extended domain of $\Gz = 56$ has confirmed their convergence during these preliminary tests.

%------------------------------------------------------------------------------------------
\section{Convergence of horizontal energy spectra}
\label{app:Convergence_of_horizontal_energy_spectra}

%------------------------------------------------------------------------------------------
\begin{figure}
\centering
\includegraphics[scale = 1.0]{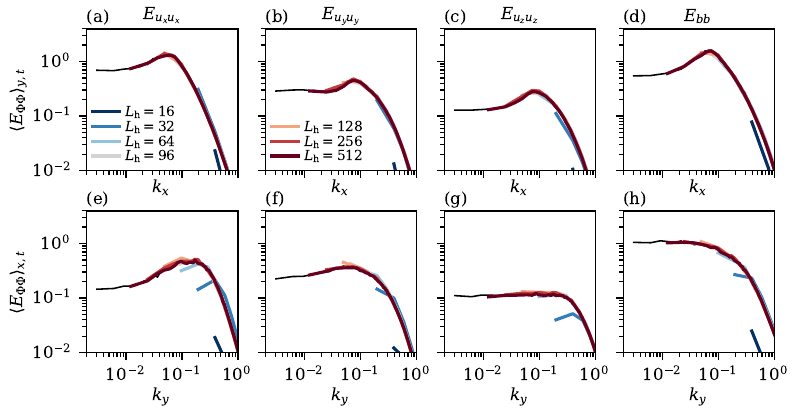}
\caption{\justifying{Convergence of energy spectra.
Both the (a -- d) streamwise and (e -- h) spanwise energy spectra converge for $\Gh \gtrsim 96$ to the accessible parts of the spectra from even larger domains. 
Thin black lines are extracted from the most extended but non-square domains ($\Gx = 2048$ and $\Gy = 512$ or vice versa).
}}
\label{fig:convergence_of_energy_spectra}
\end{figure}
%------------------------------------------------------------------------------------------

In section \ref{subsec:Anisotropy_of_emergent_large-scale_dynamics}, we have introduced Fourier energy spectra and explained that the location of their peak serves as a measure to quantify a length scale which is characteristic of the large-scale dynamics. That analysis has been conducted for our largest square domain at $\Gh = \Gamma_{\textrm{h, max}} = 512$ and underlined by even larger but non-square domains, see particularly figure \ref{fig:quantifying_the_size_of_emerging_flow_structures}.

In this appendix, via figure \ref{fig:convergence_of_energy_spectra}, we additionally contrast the Fourier energy spectra from \textit{all} our simulations with different $\Gh$.
First and foremost, we confirm that these energy spectra can indeed overlap given a sufficient $\Gh$.
We find that the energy spectra obtained from smaller domains $\Gh < \Gamma_{\textrm{h, max}}$ have indeed converged (to those from our largest domains) once $\Gh \gtrsim 96$. In contrast, smaller domains offer deviating spectra, especially along the spanwise direction as can be seen from panels (e -- h).

Let us stress that convergence of spectra means that these spectra overlap at mutually accessible wave numbers. Remember that this accessible range of wave numbers is limited by the smallest positive wave number $k_{\textrm{min}} = 2 \pi / \Gh$ due to the domain size $\Gh$. So for instance, although the spectrum of $E_{u_{x} u_{x}}$ from both $\Gh = \left\{ 96, 512 \right\}$ overlaps for $k \geq k_{\textrm{min}} \left( \Gh = 96 \right) \approx 0.07$, there is no spectrum accessible at $k < k_{\textrm{min}} \left( \Gh = 96 \right)$ for a domain of $\Gh = 96$.

This convergence has important implications. Even if a domain is too small to resolve entirely the spectral peaks $\hat{\lambda}$ forming at extreme $\Gh$, see again table \ref{tab:emergent_dynamical_scales}, the accessible part of the spectrum has still converged. In other words, the dynamics at smaller scales is then no longer affected by the missing part of the spectrum or dynamics.

%------------------------------------------------------------------------------------------
% The following will create a bibliography using the .bib file. 
\bibliographystyle{jfm}
\bibliography{jfm}

\begin{thebibliography}{53}
\expandafter\ifx\csname natexlab\endcsname\relax\def\natexlab#1{#1}\fi
\def\au#1{#1} \def\ed#1{#1} \def\yr#1{#1}\def\at#1{#1}\def\jt#1{\textit{#1}}
  \def\bt#1{#1}\def\bvol#1{\textbf{#1}} \def\vol#1{#1} \def\pg#1{#1}
  \def\publ#1{#1}\def\arxiv#1{#1}\def\org#1{#1}\def\st#1{\textit{#1}}

\bibitem[Brethouwer {\em et~al.\/}(2007)Brethouwer, Billant, Lindborg \&
  Chomaz]{brethouwer2007scaling}
{\sc \au{Brethouwer, G.}, \au{Billant, P.}, \au{Lindborg, E.} \& \au{Chomaz,
  {J.-M}}} \yr{2007}  \at{Scaling analysis and simulation of strongly
  stratified turbulent flows}.  \jt{J.~Fluid Mech.}  \bvol{585},
  \pg{343--368}.

\bibitem[Castro {\em et~al.\/}(2022)Castro, Pena, Nogueira, Gilcoto, Broullon,
  Comesana, Bouffard, Garabato \& Mourino-Carballido]{Castro2022}
{\sc \au{Castro, B.F.}, \au{Pena, M.}, \au{Nogueira, E.}, \au{Gilcoto, M.},
  \au{Broullon, E.}, \au{Comesana, A.}, \au{Bouffard, D.}, \au{Garabato, A.
  C.~N.} \& \au{Mourino-Carballido, B.}} \yr{2022}  \at{Intense upper ocean
  mixing due to large aggregations of spawning fish}.  \jt{Nat. Geosci.}
  \bvol{15}~(1),  \pg{287--292}.

\bibitem[Caulfield(2021)]{caulfield_2021}
{\sc \au{Caulfield, C.P.}} \yr{2021}  \at{Layering, instabilities and mixing in
  turbulent stratified flow}.  \jt{Annu.~Rev.~Fluid Mech.}  \bvol{53},
  \pg{113--145}.

\bibitem[Caulfield {\em et~al.\/}(1996)Caulfield, Yoshida \&
  Peltier]{caulfield_1996}
{\sc \au{Caulfield, C.P.}, \au{Yoshida, S.} \& \au{Peltier, W.R.}} \yr{1996}
  \at{Secondary instability and three-dimensionalization in a laboratory
  accelerating shear layer with varying density diﬀerences}.  \jt{Dyn. Atmos.
  Oceans}  \bvol{23},  \pg{125--138}.

\bibitem[Caulfield(2020)]{caulfield_2020}
{\sc \au{Caulfield, C.~P.}} \yr{2020}  \at{Open questions in turbulent
  stratified mixing: Do we even know what we do not know?}  \jt{Phys. Rev.
  Fluids}  \bvol{5}~(11),  \pg{110518}.

\bibitem[Deville {\em et~al.\/}(2002)Deville, Fischer \& Mund]{Deville2002}
{\sc \au{Deville, M.~O.}, \au{Fischer, P.~F.} \& \au{Mund, E.~H.}} \yr{2002}
  {\em High-order methods for incompressible fluid flow\/}.  \publ{Cambridge
  University Press}.

\bibitem[Fernando(1991)]{fernando_1991}
{\sc \au{Fernando, H. J.~S.}} \yr{1991}  \at{Turbulent mixing in stratified
  fluids}.  \jt{Annu. Rev. Fluid Mech.}  \bvol{23}~(1),  \pg{455--493}.

\bibitem[Ferrari \& Wunsch(2009)]{wunsch_2009}
{\sc \au{Ferrari, R.} \& \au{Wunsch, C.}} \yr{2009}  \at{Ocean circulation
  kinetic energy: reservoirs, sources, and sinks}.  \jt{Annu.~Rev.~Fluid Mech.}
   \bvol{4},  \pg{253--282}.

\bibitem[Fischer {\em et~al.\/}(2022)Fischer, Kerkemeier, Min, Lan, Phillips,
  Rathnayake, Merzari, Tomboulides, Karakus, Chalmers \&
  Warburton]{Fischer2022}
{\sc \au{Fischer, P.}, \au{Kerkemeier, S.}, \au{Min, M.}, \au{Lan, Y.-H.},
  \au{Phillips, M.}, \au{Rathnayake, T.}, \au{Merzari, E.}, \au{Tomboulides,
  A.}, \au{Karakus, A.}, \au{Chalmers, N.} \& \au{Warburton, T.}} \yr{2022}
  \at{Nekrs, a gpu-accelerated spectral element navier–stokes solver}.
  \jt{Parallel Comput.}  \bvol{114},  \pg{102982}.

\bibitem[Fischer(1997)]{Fischer1997}
{\sc \au{Fischer, P.~F.}} \yr{1997}  \at{An overlapping {S}chwarz method for
  spectral element solution of the incompressible navier–stokes equations}.
  \jt{J. Comput. Phys.}  \bvol{133},  \pg{84--101}.

\bibitem[Fritts {\em et~al.\/}(2022{\natexlab{{\em a\/}}})Fritts, Lund \&
  Thorpe]{fritts_2022a}
{\sc \au{Fritts, D.C.}, \au{Lund, T.S.} \& \au{Thorpe, S.A.}}
  \yr{2022{\natexlab{{\em a\/}}}}  \at{Multi-scale dynamics of
  {K}elvin–{H}elmholtz instabilities. part 1. secondary instabilities and the
  dynamics of tubes and knots}.  \jt{J.~Fluid Mech.}  \bvol{941},  \pg{A30}.

\bibitem[Fritts {\em et~al.\/}(2022{\natexlab{{\em b\/}}})Fritts, Wang, Thorpe
  \& Lund]{fritts_2022b}
{\sc \au{Fritts, D.C.}, \au{Wang, L.}, \au{Thorpe, S.A.} \& \au{Lund, T.S.}}
  \yr{2022{\natexlab{{\em b\/}}}}  \at{Multi-scale dynamics of
  {K}elvin–{H}elmholtz instabilities. part 2. energy dissipation rates,
  evolutions and statistics}.  \jt{J.~Fluid Mech.}  \bvol{941},  \pg{A31}.

\bibitem[Gibson(1987)]{gibson1987fossil}
{\sc \au{Gibson, C.H.}} \yr{1987}  \at{Fossil turbulence and intermittency in
  sampling oceanic mixing processes}.  \jt{J. Geophys. Res. Ocean.}
  \bvol{92}~(C5),  \pg{5383--5404}.

\bibitem[Gregg {\em et~al.\/}(2018)Gregg, D’Asaro, Riley \&
  Kunze]{gregg_2018}
{\sc \au{Gregg, M.C.}, \au{D’Asaro, E.A.}, \au{Riley, J.J.} \& \au{Kunze,
  E.}} \yr{2018}  \at{Mixing efficiency in the ocean}.  \jt{Annu.~Rev.~Marine
  Sci.}  \bvol{10},  \pg{443--473}.

\bibitem[Holleman {\em et~al.\/}(2016)Holleman, Geyer \&
  Ralston]{holleman_2016}
{\sc \au{Holleman, R.C.}, \au{Geyer, W.R.} \& \au{Ralston, D.K.}} \yr{2016}
  \at{Stratified turbulence and mixing efficiency in a salt wedge estuary}.
  \jt{J.~Phys.~Oceanogr.}  \bvol{46},  \pg{1769--1783}.

\bibitem[Holmboe(1962)]{Holmboe62}
{\sc \au{Holmboe, J.}} \yr{1962}  \at{On the behaviour of symmetric waves in
  stratified shear layers}.  \jt{Geophys.~Publ.}  \bvol{24},  \pg{67--113}.

\bibitem[Howard(1961)]{Howard1961}
{\sc \au{Howard, L.N.}} \yr{1961}  \at{Note on a paper of {John W. Miles}}.
  \jt{J.~Fluid Mech.}  \bvol{10},  \pg{509--512}.

\bibitem[Ivey {\em et~al.\/}(2008)Ivey, Winters \& Koseff]{ivey_2008}
{\sc \au{Ivey, G.N.}, \au{Winters, K.B.} \& \au{Koseff, J.R.}} \yr{2008}
  \at{Density stratification, turbulence, but how much mixing?}
  \jt{Annu.~Rev.~Fluid Mech.}  \bvol{40},  \pg{169--184}.

\bibitem[Kaminski \& Smyth(2019)]{Kaminski19}
{\sc \au{Kaminski, A.K.} \& \au{Smyth, W.D.}} \yr{2019}  \at{Stratified shear
  instability in a field of pre-existing turbulence}.  \jt{J.~Fluid Mech.}
  \bvol{862},  \pg{639--658}.

\bibitem[Krug {\em et~al.\/}(2020)Krug, D. \& Stevens]{Krug2020}
{\sc \au{Krug, D.}, \au{D., Lohse} \& \au{Stevens, R. J. A.~M.}} \yr{2020}
  \at{Coherence of temperature and velocity superstructures in turbulent
  {R}ayleigh-{B}\'enard flow}.  \jt{J. Fluid Mech.}  \bvol{887},  \pg{A2}.

\bibitem[Laurent {\em et~al.\/}(2002)Laurent, Simmons \& Jayne]{StLaurent2002}
{\sc \au{Laurent, L.C.~St.}, \au{Simmons, H.~L.} \& \au{Jayne, S.R.}} \yr{2002}
   \at{Estimating tidally driven mixing in the deep ocean}.  \jt{Geophys. Res.
  Lett.}  \bvol{29},  \pg{21--1--21--4}.

\bibitem[Liu {\em et~al.\/}(2022)Liu, Kaminski \& Smyth]{liu_2022}
{\sc \au{Liu, C.-L.}, \au{Kaminski, A.K.} \& \au{Smyth, W.D.}} \yr{2022}
  \at{The butterfly effect and the transition to turbulence in a stratified
  shear layer}.  \jt{J.~Fluid Mech.}  \bvol{953},  \pg{A43}.

\bibitem[Lorenz(1955)]{lorenz_1955}
{\sc \au{Lorenz, E.~N.}} \yr{1955}  \at{Available potential energy and the
  maintenance of the general circulation}.  \jt{Tellus}  \bvol{7},
  \pg{157--167}.

\bibitem[Mashayek \& Peltier(2012{\natexlab{{\em a\/}}})]{mashayek_2012a}
{\sc \au{Mashayek, A.} \& \au{Peltier, W.R.}} \yr{2012{\natexlab{{\em a\/}}}}
  \at{The `zoo' of secondary instabilities precursory to stratified shear flow
  transition. {Part 1}: Shear aligned convection, pairing, and braid
  instabilities}.  \jt{J.~Fluid Mech.}  \bvol{708},  \pg{5--44}.

\bibitem[Mashayek \& Peltier(2012{\natexlab{{\em b\/}}})]{mashayek_2012b}
{\sc \au{Mashayek, A.} \& \au{Peltier, W.R.}} \yr{2012{\natexlab{{\em b\/}}}}
  \at{The `zoo' of secondary instabilities precursory to stratified shear flow
  transition. {Part 2}: The influence of stratification}.  \jt{J.~Fluid Mech.}
  \bvol{708},  \pg{45--70}.

\bibitem[Mashayek \& Peltier(2013)]{mashayek_2013}
{\sc \au{Mashayek, A.} \& \au{Peltier, W.R.}} \yr{2013}  \at{Shear-induced
  mixing in geophysical flows: does the route to turbulence matter to its
  efficiency?}  \jt{J.~Fluid Mech.}  \bvol{725},  \pg{216--261}.

\bibitem[Mater \& Venayagamoorthy(2014)]{Mater2014}
{\sc \au{Mater, Benjamin~D.} \& \au{Venayagamoorthy, Subhas~Karan}} \yr{2014}
  \at{A unifying framework for parameterizing stably stratified shear-flow
  turbulence}.  \jt{Phys. Fluids}  \bvol{26}~(3),  \pg{036601}.

\bibitem[Miles(1961)]{Miles1961}
{\sc \au{Miles, J.W.}} \yr{1961}  \at{On the instability of heterogeneous shear
  flows}.  \jt{J.~Fluid Mech.}  \bvol{10},  \pg{496--508}.

\bibitem[Osborn(1980)]{osborn_1980}
{\sc \au{Osborn, T.R.}} \yr{1980}  \at{Estimates of the local rate of vertical
  diffusion from dissipation measurements}.  \jt{J.~Phys.~Oceanogr.}
  \bvol{10},  \pg{83--89}.

\bibitem[Peltier \& Caulfield(2003)]{peltier_2003}
{\sc \au{Peltier, W.R.} \& \au{Caulfield, C.P.}} \yr{2003}  \at{Mixing
  efficiency in stratified shear flows}.  \jt{Annu. Rev. Fluid Mech.}
  \bvol{35}~(1),  \pg{135--167}.

\bibitem[Portwood {\em et~al.\/}(2019)Portwood, de~Bruyn~Kops \&
  Caulfield]{portwood_2019}
{\sc \au{Portwood, G.D.}, \au{de~Bruyn~Kops, S.M.} \& \au{Caulfield, C.P.}}
  \yr{2019}  \at{Asymptotic dynamics of high dynamic range stratified
  turbulence}.  \jt{Phys.~ Rev.~ Letters}  \bvol{122}~(19),  \pg{194504}.

\bibitem[Salehipour {\em et~al.\/}(2016)Salehipour, Caulfield \&
  Peltier]{salehipour_2016}
{\sc \au{Salehipour, H.}, \au{Caulfield, C.P.} \& \au{Peltier, W.R.}} \yr{2016}
   \at{Turbulent mixing due to the {Holmboe} wave instability at high reynolds
  number}.  \jt{J.~Fluid Mech.}  \bvol{803},  \pg{591--621}.

\bibitem[Salehipour {\em et~al.\/}(2018)Salehipour, Peltier \&
  Caulfield]{salehipour_2018}
{\sc \au{Salehipour, H.}, \au{Peltier, W.R.} \& \au{Caulfield, C.P.}} \yr{2018}
   \at{Self-organized criticality of turbulence in strongly stratified mixing
  layers}.  \jt{J.~Fluid Mech.}  \bvol{856},  \pg{228--256}.

\bibitem[Scheel {\em et~al.\/}(2013)Scheel, Emran \& Schumacher]{Scheel2013}
{\sc \au{Scheel, J.~D.}, \au{Emran, M.~S.} \& \au{Schumacher, J.}} \yr{2013}
  \at{An overlapping schwarz method for spectral element solution of the
  incompressible {N}avier–{S}tokes equations}.  \jt{New J. Phys.}  \bvol{15},
   \pg{113063}.

\bibitem[Scinocca \& Ford(2000)]{scinocca_2000}
{\sc \au{Scinocca, J.F.} \& \au{Ford, R.}} \yr{2000}  \at{The nonlinear forcing
  of large-scale internal gravity waves by stratified shear instability}.
  \jt{J. Atmosp. Sci.}  \bvol{57},  \pg{653--672}.

\bibitem[Shih {\em et~al.\/}(2005)Shih, Koseff, Ivey \& Ferziger]{Shih05}
{\sc \au{Shih, L.H.}, \au{Koseff, J.R.}, \au{Ivey, G.N.} \& \au{Ferziger,
  J.H.}} \yr{2005}  \at{Parameterization of turbulent fluxes and scales using
  homogeneous sheared stably stratified turbulence simulations}.  \jt{J.~Fluid
  Mech.}  \bvol{525},  \pg{193--214}.

\bibitem[Smith {\em et~al.\/}(2021)Smith, Caulfield \& Taylor]{Smith2021}
{\sc \au{Smith, K.~M.}, \au{Caulfield, C.~P.} \& \au{Taylor, J.~R.}} \yr{2021}
  \at{Turbulence in forced stratified shear flows}.  \jt{J.~Fluid Mech.}
  \bvol{910},  \pg{A42}.

\bibitem[Smyth \& Moum(2000)]{Smyth2000}
{\sc \au{Smyth, W.D.} \& \au{Moum, J.N.}} \yr{2000}  \at{Length scales of
  turbulence in stably stratified mixing layers}.  \jt{Phys.~Fluids}
  \bvol{12},  \pg{1327--1342}.

\bibitem[Smyth {\em et~al.\/}(2011)Smyth, Moum \& Nash]{Smyth11}
{\sc \au{Smyth, W.D.}, \au{Moum, J.N.} \& \au{Nash, J.D.}} \yr{2011}
  \at{Narrowband, high-frequency oscillations at the equator. {Part II}:
  {Properties} of shear instabilities}.  \jt{J.~Phys.~Oceanogr.}  \bvol{41},
  \pg{412--428}.

\bibitem[Smyth {\em et~al.\/}(2019)Smyth, Nash \& Moum]{smyth_2019}
{\sc \au{Smyth, W.D.}, \au{Nash, J.~D.} \& \au{Moum, J.~N.}} \yr{2019}
  \at{Self-organized criticality in geophysical turbulence}.  \jt{Sci. Rep.}
  \bvol{9},  \pg{3747}.

\bibitem[Smyth {\em et~al.\/}(1988)Smyth, Klaassen \& Peltier]{smyth_1988}
{\sc \au{Smyth, W.~D.}, \au{Klaassen, G.~P.} \& \au{Peltier, W.~R.}} \yr{1988}
  \at{Finite amplitude holmboe waves}.  \jt{Geophys. Astrophys. Fluid Dyn.}
  \bvol{43},  \pg{181--222}.

\bibitem[Taylor \& Thompson(2023)]{taylor_2023}
{\sc \au{Taylor, J.~R.} \& \au{Thompson, A.~F.}} \yr{2023}  \at{Submesoscale
  dynamics in the upper ocean}.  \jt{Annu.~Rev.~Fluid Mech.}  \bvol{55},
  \pg{103--127}.

\bibitem[Thorpe(1985)]{thorpe_1985}
{\sc \au{Thorpe, S.A.}} \yr{1985}  \at{Laboratory observations of secondary
  structures in {K}elvin–{H}elmholtz billows and consequences for ocean
  mixing}.  \jt{Geophys. Astrophys. Fluid Dyn.}  \bvol{34},  \pg{175--199}.

\bibitem[Thorpe(1987)]{thorpe_1987}
{\sc \au{Thorpe, S.A.}} \yr{1987}  \at{Transitional phenomena and the
  development of turbulence in stratified fluids: a review}.  \jt{J. Geophys.
  Res.}  \bvol{92},  \pg{5231–--5248}.

\bibitem[Thorpe(2002)]{thorpe_2002}
{\sc \au{Thorpe, S.A.}} \yr{2002}  \at{The axial coherence of
  {K}elvin–{H}elmholtz billows}.  \jt{Q. J. R. Meteorol. Soc.}  \bvol{128},
  \pg{1529–--1542}.

\bibitem[Thorpe(2005)]{Thorpe2005}
{\sc \au{Thorpe, S.A.}} \yr{2005} {\em The Turbulent Ocean\/}.  \publ{Cambridge
  University Press}.

\bibitem[Turner(1973)]{turner_1973}
{\sc \au{Turner, J.~S.}} \yr{1973} {\em Buoyancy effects in fluids\/}.
  \publ{Cambridge University Press}.

\bibitem[Uncles \& Mitchell(2011)]{Uncles2011}
{\sc \au{Uncles, R.~J.} \& \au{Mitchell, S.~B.}} \yr{2011}  \at{Turbidity in
  the {T}hames estuary: How turbid do we expect it to be?}  \jt{Hydrobiologia}
  \bvol{672},  \pg{91--103}.

\bibitem[Vieweg(2023)]{Vieweg2023a}
{\sc \au{Vieweg, P.P.}} \yr{2023}  \at{Large-scale flow structures in turbulent
  {R}ayleigh-{B}énard convection: Dynamical origin, formation, and role in
  material transport}. PhD thesis, TU Ilmenau.

\bibitem[Watanabe \& Nagata(2021)]{Watanabe2021}
{\sc \au{Watanabe, Tomoaki} \& \au{Nagata, Koji}} \yr{2021}  \at{Large-scale
  characteristics of a stably stratified turbulent shear layer}.  \jt{J. Fluid
  Mech.}  \bvol{927},  \pg{A27}.

\bibitem[Winters {\em et~al.\/}(1995)Winters, Lombard, Riley \&
  D'Asaro]{Winters95}
{\sc \au{Winters, K.B.}, \au{Lombard, P.N.}, \au{Riley, J.J.} \& \au{D'Asaro,
  E.A.}} \yr{1995}  \at{Available potential energy and mixing in
  density-stratified fluids}.  \jt{J.~Fluid Mech.}  \bvol{289},  \pg{115--128}.

\bibitem[Wunsch \& Ferrari(2004)]{wunsch_2004}
{\sc \au{Wunsch, C.} \& \au{Ferrari, R.}} \yr{2004}  \at{Vertical mixing,
  energy and the general circulation of the oceans}.  \jt{Annu.~Rev.~Fluid
  Mech.}  \bvol{36},  \pg{281--314}.

\bibitem[Zhou {\em et~al.\/}(2017)Zhou, Taylor \& Caulfield]{zhou_2017}
{\sc \au{Zhou, Q.}, \au{Taylor, J.R.} \& \au{Caulfield, C.P.}} \yr{2017}
  \at{Self-similar mixing in stratified plane couette flow for varying prandtl
  number}.  \jt{J.~Fluid Mech.}  \bvol{820},  \pg{86--120}.

\end{thebibliography}

%------------------------------------------------------------------------------------------
\end{document}